\documentclass{llncs}
\usepackage{graphicx}
\usepackage{amsmath}
\usepackage{amssymb}
\usepackage{enumerate}
\usepackage{varioref}

\def\boxit#1{\vbox{\hrule\hbox{\vrule\kern4pt
  \vbox{\kern1pt#1\kern1pt}
\kern2pt\vrule}\hrule}}

\usepackage{url}

\newcommand\nc{\newcommand}

\newtheorem{rules}{Rule}

\nc{\crl}[2]{\begin{corollary}\label{crl:#1} #2 \end{corollary}}
\nc{\dfn}[2]{\begin{definition}\label{def:#1} #2 \end{definition}}
\nc{\lem}[2]{\begin{lemma}\label{lem:#1} #2 \end{lemma}}
\nc{\prp}[2]{\begin{proposition}\label{prp:#1} #2
\end{proposition}}
\nc{\thm}[2]{\begin{theorem}\label{thm:#1} #2\end{theorem}}
\nc{\fac}[2]{\begin{lemma}\label{fact:#1} #2 \end{lemma}}
\nc{\rul}[2]{\begin{rules}\label{rul:#1} #2 \end{rules}}

\nc{\eqn}[2]{\begin{eqnarray}\label{eqn:#1} #2 \end{eqnarray}}

\nc{\fig}[4]{\begin{figure}[h]
\begin{center}
\includegraphics[width=#2\textwidth]{#4}
\end{center}
\caption{#3}\label{fig:#1}
\end{figure}}

\nc{\tbl}[3]{\begin{table}[hbt] #3 \caption{#2} \label{tab:#1}
\end{table}}

\nc{\refc}[1]{Corollary~\ref{crl:#1}}
\nc{\refd}[1]{Definition~\ref{def:#1}}
\nc{\reff}[1]{Figure~\ref{fig:#1}}
\nc{\refl}[1]{Lemma~\ref{lem:#1}}
\nc{\refp}[1]{Proposition~\ref{prp:#1}}
\nc{\reft}[1]{Theorem~\ref{thm:#1}} \nc{\refe}[1]{(\ref{eqn:#1})}
\nc{\reftb}[1]{Table~\ref{tab:#1}}

\nc{\reffc}[1]{Fact~\ref{fact:#1}}
\nc{\refr}[1]{Rule~\ref{rul:#1}}

\nc{\pf}[1]{ \noindent \emph{Proof.} #1 \hfill \qed\par}

\long\def\invis#1{}

\title{Characterizing Star-PCGs
}

\author{Mingyu Xiao\inst{1}
\and
Hiroshi Nagamochi\inst{2}}

 \institute{
 School of Computer Science and Engineering,
University of Electronic Science and Technology of China, China,
 \email{myxiao@gmail.com}
 \and
 Department of Applied Mathematics and Physics,
  Graduate School of Informatics, Kyoto University, Japan,
 \email{nag@amp.i.kyoto-u.ac.jp}}

\begin{document}

\maketitle

\begin{abstract}
%
A graph $G$ is called a  pairwise compatibility graph  (PCG, for short)
if it admits a tuple $(T,w, d_{\min},d_{\max})$ of
  a tree $T$ whose leaf set is equal to the vertex set of $G$,
a non-negative edge weight $w$,
and two non-negative reals $d_{\min}\leq d_{\max}$
 such that $G$ has an edge between  two vertices $u,v\in V$  if and only if
the distance between the two leaves $u$ and $v$ in the weighted tree $(T,w)$ is
in the interval $[d_{\min}, d_{\max}]$.
 The tree $T$ is also called a witness tree of the PCG $G$.
The problem of testing if a given graph is a PCG is not known to be NP-hard yet.
To obtain a complete characterization of PCGs is a wide open  problem
 in computational biology and graph theory.
In literature, most witness trees
admitted by known PCGs are stars and caterpillars.
In this paper, we give a complete characterization for a graph to be a star-PCG
(a PCG that admits a star as its witness tree),   which provides
us  the first polynomial-time algorithm for recognizing star-PCGs.

 \vspace*{5mm} \noindent {\bf Key words.} \ \
Pairwise Compatibility Graph;
Polynomial-time Algorithm;
Graph Algorithm;
Graph Theory
\end{abstract}


\section{Introduction}\label{sec:intro}

Pairwise compatibility graph is a graph class originally motivated from computational biology.
In biology, the evolutionary history of a set of organisms is represented by a phylogenetic tree, which
is a tree with leaves representing known taxa and internal nodes representing ancestors
 that might have led to these taxa through evolution.
Moreover, the edges in the phylogenetic tree may be assigned weights to represent the evolutionary distance among species.
Given a set of taxa and some relations among the taxa,
we may want to construct a phylogenetic tree of the taxa.
The set of taxa may be a subset of taxa from a large phylogenetic tree, subject to some biologically-motivated constraints.
Kearney, Munro and Phillips~\cite{phtree} considered the following constraint on
sampling based on the observation in~\cite{phtree0}:
the pairwise distance between any two leaves in the sample phylogenetic tree is
between two given integers $d_{min}$ and $d_{max}$.
This motivates the introduction of pairwise compatibility graphs (PCGs).
Given a phylogenetic tree $T$ with an edge weight $w$ and two real numbers
 $d_{min}$ and $d_{max}$, we can construct a graph $G$ each vertex of which is corresponding
to a leaf of $T$ 
so that
there is an edge between two vertices in $G$ if and only if
 the corresponding two leaves of $T$ are at a distance within the interval $[d_{min},d_{max}]$ in $T$.
The graph $G$ is called the PCG of the tuple $(T,w, d_{min},d_{max})$.
Nowadays, PCG becomes an interesting graph class and topic in graph theory.
Plenty of structural results have been developed.

It is 
straightforward
 to construct a PCG from a given tuple  $(T,w, d_{min},d_{max})$.
However, the inverse direction seems a considerably hard task.
Few methods have been known for constructing
 a corresponding tuple $(T,w, d_{min},d_{max})$ from a given graph $G$.
The inverse problem attracts certain interests in graph algorithms, which may also have potential applications in computational biology.
It has been extensively studied from many aspects after the introduction of
PCG~\cite{CFS2013,splitgraph,survey,noPCG,Y-PCG,DiscoveringPCG}.

A natural question 
was  whether all graphs are PCGs.
This was  proposed as a conjecture 
 in~\cite{phtree}, and was confuted in~\cite{DiscoveringPCG}
by giving a counterexample  
of a bipartite graph with with 15 vertices.
Later, a counterexample with
eight vertices and a counterexample of a planar graph with 20 vertices
were found~\cite{noPCG}.
It has been checked that all graphs with at most
seven vertices are PCGs~\cite{CFS2013}
and all bipartite graphs with at most
eight vertices are PCGs~\cite{revisited}.
In fact, it is even not easy to check whether a graph with a small constant number of vertices is a PCG or not.
%
Whether recognizing PCGs is NP-hard or not is currently open.
Some references conjecture the NP-hardness of the problem~\cite{survey,noPCG}.
 A generalized version of PCG recognition is shown to be NP-hard~\cite{noPCG}.

PCG also becomes an interesting graph class in graph theory.
It contains the well-studied graph class of \emph{leaf power graphs} (LPGs) as a subset of instances
such that $d_{min}=0$,
which was introduced in the context of constructing phylogenies from species similarity data~\cite{LPG1,LPG2,LPG3}.
Another natural relaxation of PCG is to set $d_{max}=\infty$. This graph class is known as \emph{min leaf power graph} (mLPG)~\cite{splitgraph}, which is the complement of LPG.
Several other known graph classes have been shown to be 
subclasses of PCG,
e.g., disjoint union of cliques~\cite{leafpowers}, forests~\cite{conditions}, chordless cycles and single
chord cycles~\cite{Y-PCG}, tree power graphs~\cite{DiscoveringPCG},  threshold graphs~\cite{splitgraph}, triangle-free outerplanar 3-graphs~\cite{triang-free}, some particular
subclasses of split matrogenic graphs~\cite{splitgraph}, Dilworth 2 graphs~\cite{dilworth2}, the complement of a forest~\cite{conditions} and so on.
It is also known that a PCG with a witness tree being a caterpillar also allows a witness tree being a centipede~\cite{C2013}.
A method for constructing PCGs  is derived \cite{reductionXN},
where it is shown that
a graph $G$ consisting two graphs $G_1$ and $G_2$ that share a vertex as a cut-vertex in $G$
is a PCG if and only both $G_1$ and $G_2$ are PCGs.

How to recognize PCGs or construct a corresponding phylogenetic tree for a PCG
has become an interesting open problem in this area.
To make a step toward this open problem, we consider PCGs with a witness tree being a star in this paper,
which 
we call star-PCGs.
One motivation why we consider stars is that: in the literature,
most of the witness trees of PCGs have simple graph structures, such as stars and caterpillars~\cite{survey}.
It 
is also fundamental
to consider the problem of characterizing subclasses of PCGs
derived from a specific topology of trees.
Although stars are trees with a rather simple topology, star-PCG recognition is not easy at all.
It is known that threshold graphs  are star-PCGs (even in star-LPG and star-mLPG) and
the class of star-PCGs is nearly the class of three-threshold graphs,
a graph class extended from the threshold graphs~\cite{splitgraph}.
However, no complete characterization of star-PCGs and no polynomial-time recognition of star-PCGs
are known.
In this paper, we give a complete characterization for a graph to be a star-PCG,
which provides
us  the first polynomial-time algorithm for recognizing star-PCGs.

The main idea of our algorithm is as follows.
Without loss of generality, we always rank the leaves of the witness star $T_V$
 (and the corresponding vertices in the star-PCG $G$) according to the weight of the edges incident on it.
When such an ordering of the vertices in a  star-PCG $G$ is given,
we can see that
all the neighbors of each vertex in $G$ must appear consecutively in the  ordering.
This motivates us to define 
such an ordering to be
 ``consecutive ordering.''
 To check if a graph is a star-PCG, we can first check if the graph
can have a consecutive ordering of vertices.
Consecutive orderings can be computed in polynomial time by reducing to
the problem of recognizing interval graphs.
However, this is not enough to test star-PCGs.
A graph may not be a star-PCG even if it has a consecutive ordering of vertices.
We further investigate the structural properties of star-PCGs
on a fixed
 consecutive ordering of vertices.
We find that three cases of non-adjacent vertex pairs, called \emph{gaps},
can be used to characterize star-PCGs.
A graph is a star-PCG if and only if 
it admits
a consecutive ordering of vertices that is gap-free (Theorem~\ref{thm:order}).
Finally, to show that whether a given graph is gap-free or not can be tested in polynomial time
 (Theorem~\ref{th:test_gap-free}), we also use
a notion of
``contiguous orderings.''
All these together contribute to a polynomial-time algorithm for our problem.

 The paper is organized as follows.
 Section~\ref{sec:prelimi} introduces some basic notions and notations necessary to this paper.
 Section~\ref{sec:family} discusses how to  test  whether a given family  $\mathcal{S}$
 of subsets of an element set $V$ admits
  a special  ordering on $V$, called
  ``consecutive'' or ``contiguous'' orderings and
  proves the uniqueness of such orderings under some conditions on  $\mathcal{S}$.
  This uniqueness plays a key role to prove that
  whether a given graph is a star-PCG or not can tested in polynomial time.
     Section~\ref {sec:weight} characterizes the class of star-PCGs $G=(V,E)$ in terms of
    an ordering $\sigma$ of the vertex set $V$, called a  ``gap-free'' ordering,
    and shows that given  a gap-free  ordering of $V$, a tuple
    $(T,w,d_{\min},d_{\max})$ that represents  $G$   can be computed in polynomial time.
      Section~\ref{sec:recognize} first derives structural properties on a graph that admits
    a  ``gap-free'' ordering, and then presents a method for testing if a given graph is a star-PCG or not
    in polynomial time by using the result on contiguous orderings to a family of sets.
    Finally
    Section~\ref{sec:conclude} makes some concluding remarks.
    Due to the space limitation, some proofs are moved to Appendix.

\section{Preliminaries}\label{sec:prelimi}

For two integers $a$ and $b$, let $[a,b]$ denote the set of integers $i$ with $a\leq i\leq b$.
For a sequence $\sigma$ of elements, let $\overline{\sigma}$ denote the reversal of $\sigma$.
A sequence obtained by concatenating two sequences $\sigma_1$ and $\sigma_2$ in this order
is denoted by $(\sigma_1, \sigma_2)$.

\paragraph{\textbf{Families of Sets}.}
Let $V$ be a set of $n\geq 1$ elements.
We call a subset $S\in V$ {\em trivial} in $V$ if $|S|\leq 1$ or $S=V$.
We say that a set $X$ has a common element with a set $Y$ if
$X\cap Y\neq\emptyset$.
 We say that two subsets $X,Y\subseteq V$ {\em intersect}
 (or $X$ intersects $Y$) if
 three sets
  $X\cap Y$, $X\setminus Y$, and $Y\setminus X$  are
 all
 non-empty sets.
A {\em partition} $\{V_1,V_2,\ldots,V_k\}$ of $V$ is defined to be
a collection of disjoint non-empty subsets $V_i$ of $V$ such that
their union is $V$, where possibly $k=1$.

Let
  $\mathcal{S}\subseteq 2^V$ be a family of $m$ subsets of 
  $V$.
A total ordering $u_1,u_2,\ldots,u_n$ of elements in $V$ is called
{\em consecutive} to  $\mathcal{S}$  if
each non-empty set $S\in \mathcal{S}$ consists of elements with consecutive indices,
i.e.,
$S$ is equal to $\{u_i,u_{i+1},\ldots, u_{i+|S|-1}\}$ for some $i\in [1, n-|S|-1]$.
A   consecutive  ordering $u_1,u_2,\ldots,u_n$ of elements in $V$ to  $\mathcal{S}$ is called
 {\em contiguous} if
 any two sets $S,S'\in \mathcal{S}$ with $S'\subseteq S$ start from or end with the same element
 along the ordering, i.e.,
   $S'=\{u_j,u_{j+1},\ldots, u_{j+|S'|-1}\}$ and $S=\{u_i,u_{i+1},\ldots, u_{i+|S|-1}\}$
  satisfy  $j=i$ or $j+|S'|  =i+|S| $.

\paragraph{\textbf{Graphs}.} 
Let a graph  stand for a simple undirected graph. 
A graph (resp., bipartite graph) with a vertex set $V$ and an edge set $E$
(resp., an edge set $E$ between two vertex sets $V_1$ and $V_2=V\setminus V_1$)
is denoted by $G=(V,E)$ (resp.,   $(V_1,V_2,E)$).
Let $G$ be a graph, where  $V(G)$ and $E(G)$ denote  the sets of vertices and edges
in a graph $G$, respectively.
For a vertex $v$ in $G$,
we denote by  $N_G(v)$ the set of neighbors of a vertex $v$ in $G$,
and define {\em degree} $\mathrm{deg}_G(v)$ to be the  $|N_G(v)|$.
We call a pair of vertices $u$ and $v$ in  $G$
a {\em mirror pair} if $N_G(v)\setminus \{u\}=N_G(u)\setminus \{v\}$.
Let $X$ be a subset of $V(G)$.
Define $N_G(X)$ to be the set 
of neighbors of $X$, i.e., $N_G(X)=\{u\in N_G(v)\setminus X\mid v\in X\}$.
Let $G-X$ denote
the graph obtained from $G$ by removing vertices in $X$
together with all edges incident to vertices in $X$,
where $G-\{v\}$ for a vertex $v$ may be written as $G-v$.
Let $G[X]$ denote the graph induced by
$X$,
i.e., $G[X]= G -(V(G)\setminus X)$.

Let $T$ be a tree.
A vertex $v$ in $T$ is called an {\em inner vertex} if $\mathrm{dege}_T(v)\geq 2$
and is called a {\em leaf} otherwise.
Let $L(T)$ denote the set of leaves.
An edge incident to a leaf in $T$ is called a {\em leaf edge} of $T$.
A tree $T$ is called a {\em star} if it has at most one inner vertex.

\paragraph{\textbf{Weighted Graphs}.}
An edge-weighted graph $(G,w)$ is defined to be a pair of
a graph $G$ and a non-negative weight function $w:E(G) \to \Re_+$.
 For a subgraph $G'$ of $G$, let $w(G')$ denote
 the sum $\sum_{e\in E(G')}w(e)$ of edge weights in $G'$.

Let $(T,w)$ be an edge-weighted tree.
For two vertices $u,v\in V(T)$,
let $\mathrm{d}_{T,w}(u,v)$ denote
  the sum of weights of edges in the 
  unique
   path  of $T$ between $u$ and $v$.

\paragraph{\textbf{PCGs}.}
For a tuple $(T,w,d_{\min}, d_{\max})$ of an edge-weighted tree $(T,w)$ and
two  non-negative reals $d_{\min}$ and $d_{\max}$,
define $G(T,w,d_{\min}, d_{\max})$ to be the simple graph
$(L(T), E)$ such that, for any two distinct vertices $u,v\in L(T)$,
$uv\in E$ if and only if $d_{\min} \leq \mathrm{d}_{T,w}(u,v)\leq d_{\max}$.
Note that $G(T,w,d_{\min}, d_{\max})$ is not necessarily connected.

\begin{figure}[h!] \begin{center}
 \includegraphics[scale=0.15]{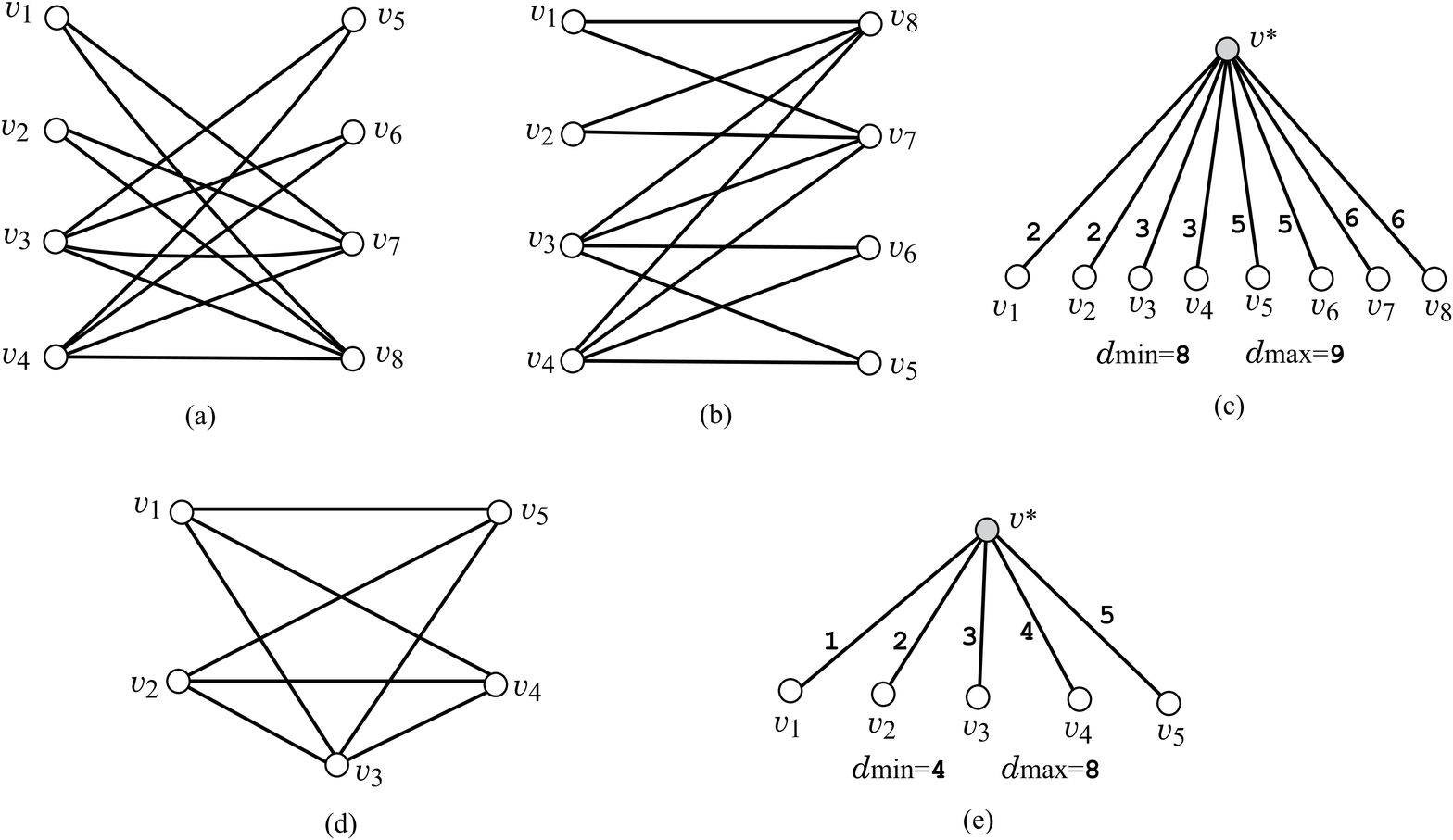} \end{center}
 \caption{Illustration of examples of star-PCG.
 (a)   A connected and  bipartite star-PCG $G_1=(V_1,V_2,E)$,
 where  ordering  $\sigma_a=v_1,v_2,v_3,v_4,v_8,v_7,v_6,v_5$  is not gap-free to $G_1$.
 (b)  
  $G_1$ in (a)
   with a gap-free  ordering  $\sigma_b=v_1,v_2,\ldots,v_8$.
  (c) A PCR $(T,w,d_{\min}=8, d_{\max}=9)$ of  
   $G_1$ in (b).
 %
 (d)  A connected and non-bipartite star-PCG $G_2$.
 (e) A PCR $(T,w,d_{\min}=4, d_{\max}=8)$ of  
 $G_2$ in (d).
  } \label{fi:PCG_example}
 \end{figure}

A graph $G$ is called a {\em pairwise compatibility graph} (PCG, for short)
if there exists a tuple $(T,w,d_{\min}, d_{\max})$ such that
$G$ is isomorphic to the graph $G(T,d_{\min}, d_{\max})$,
where
we call such a  tuple a   {\em pairwise compatibility representation} (PCR, for short) of $G$,
and call a tree $T$ in a PCR  of $G$ a {\em pairwise compatibility tree}
 (PCT, for short) of $G$. The tree $T$ is called a \emph{witness tree} of $G$.
 We call a PCG $G$ a {\em star-PCG} if it admits a PCR $(T,w,d_{\min}, d_{\max})$
 such that $T$ is a star.
 Fig.~\ref{fi:PCG_example} illustrates examples of star-PCGs and PCRs of them.
 Although phylogenetic trees may not have edges with weight 0 or degree-2 vertices by some biological motivations~\cite{C2013},
 our PCTs do not have these constraints.
 This 
 relaxation
 will be helpful for us to analyze structural properties of PCGs from graph theory.
 Furthermore, it is easy to get rid of edges with weight 0 or degree-2 vertices in a tree by
 contracting
 an edge.

\begin{lemma}\label{lem_ini}
Every PCG admits a PCR $(T,w,d_{\min}, d_{\max})$ such that
  $0<d_{\min}< d_{\max}$ and $w(e)>0$ 
  for all edges $e\in E(T)$.
\end{lemma}

 \vspace{-4mm}
\section{Consecutive/Contiguous Orderings of Elements}\label{sec:family}

Let
  $\mathcal{S}\subseteq 2^V$ be a family of $m$ subsets of 
  a set $V$ of $n\geq 1$ elements in this section.
  Let $V(\mathcal{S})$ denote the union of all subsets in $\mathcal{S}$,
  and $\pi(\mathcal{S})$ denote the partition  $\{V_1,V_2,\ldots,V_p\}$ of $V(\mathcal{S})$
  such that $u,v\in V_i$ for some $i$ if and only if $\mathcal{S}$ has no set $S$ with $|\{u,v\}\cap S|=1$.
An auxiliary graph $H_{\mathcal{S}}$ for $\mathcal{S}$ is
defined to be the graph $(\mathcal{S}, E_{\mathcal{S}})$
 that joins two sets $S,S'\in \mathcal{S}$ with an edge $S S'\in  E_{\mathcal{S}}$
  if and only if $S$ and $S'$ intersect.

\subsection{Consecutive  Orderings of Elements}

Observe that 
when $\mathcal{S}$ admits a consecutive ordering of $V(\mathcal{S})$,
any subfamily $\mathcal{S}'\subseteq \mathcal{S}$  admits
a consecutive ordering of $V(\mathcal{S}')$.
We call a non-trivial   set $C\subseteq V$ a {\em cut} to 
 $\mathcal{S}$
if no  set $S\in \mathcal{S}$ intersects $C$,
i.e.,  each $S\in \mathcal{S}$ satisfies one of  $S\supseteq C$, $S\subseteq C$ and  $S\cap C=\emptyset$.
We call  $\mathcal{S}$ {\em cut-free}
 if $\mathcal{S}$ has no cut.

\begin{theorem}\label{th:consecutive}
For a set $V$ of  $n\geq 1$ elements and a   family  $\mathcal{S}\subseteq 2^V$ of $m\geq 1$ sets,
  a consecutive ordering of $V$ to $\mathcal{S}$ can be found
  in $O(nm^2 )$ time, if one exists.
  Moreover
 if  $\mathcal{S}$ is   cut-free, then
 a consecutive ordering of $V$ to $\mathcal{S}$ is unique up to reversal.
\end{theorem}


\subsection{Contiguous Orderings of Elements}


 We call two elements $u,v\in V$ {\em equivalent} in  $\mathcal{S}$
  if    no set $S\in \mathcal{S}$ satisfies $|\{u,v\}\cap S|=1$.
 We call    $\mathcal{S}$
  {\em simple} if there is no pair of  equivalent  elements $u,v\in V$.
  Define $\mathcal{X}_{\mathcal{S}}$ to be
  the family of maximal sets $X\subseteq V$
  such that any two vertices in $X$ are equivalent and
  $X$ is maximal subject to this property.

A  non-trivial  set $S\in \mathcal{S}$ is called a {\em separator}
if no other set $S'\in \mathcal{S}$ contains or intersects $S$,
i.e.,  each $S'\in \mathcal{S}$ satisfies  $S'\subseteq S$ or  $S'\cap S=\emptyset$.
We call  $\mathcal{S}$  {\em separator-free} in $\mathcal{S}$
 if $\mathcal{S}$ has no    separator.

\begin{theorem}\label{th:contiguous} 
For a set $V$ of  $n\geq 1$ elements and 
a  family  $\mathcal{S}\subseteq 2^V$ of $m\geq 1$ sets,
  a contiguous ordering of $V$ to $\mathcal{S}$ can be found in 
 $O(nm^2)$  time, if one exists.
  Moreover 
 all elements in each set $X\in \mathcal{X}_{\mathcal{S}}$ appear
 consecutively in any contiguous ordering of $V$ to $\mathcal{S}$,
 and a  contiguous ordering of $V$ to $\mathcal{S}$ is unique up to reversal 
 of the entire ordering and
 arbitrariness of orderings of elements in each set $X\in \mathcal{X}_{\mathcal{S}}$.
\end{theorem}  


\section{Star-PCGs}\label{sec:weight}

Let $G=(V,E)$ be a graph with $n\geq 2$ vertices, not necessarily connected.
Let $M_G$ denote the set of mirror pairs $\{u,v\}\subseteq V$ in $G$, i.e.,
$N_G(u)\setminus \{v\}= N_G(v)\setminus \{u\}$,
where $u$ and $v$ are not necessarily adjacent.
Let $T_V$ be a star with a center $v^*$ and $L(T)=V$.
An {\em ordering}     of $V$ is defined to be a bijection
$\sigma: V\to \{1,2,\ldots,n\}$, and
we simply write a vertex $v$ with $\sigma(v)=i$ with $v_i$.
For an edge weight $w$ in $T_V$,
we simply denote $w(v^*v_i)$ by $w_i$.
When $G$ is a star-PCG of a tuple $(T_V,w,d_{\min},d_{\max})$,
there is an ordering $\sigma$ of  $V$ such that
$w_1\leq w_2\leq\cdots\leq w_n$.
Conversely this section
 derives a necessary and sufficient condition for a pair $(G,\sigma)$ of a graph $G$
and  an ordering $\sigma$ of $V$
to admit a PCR $(T_V,w,d_{\min},d_{\max})$ of $G$ such that
$w_1\leq w_2\leq\cdots\leq w_n$.

 \begin{figure}[htbp] \begin{center}
 \includegraphics[scale=0.12]{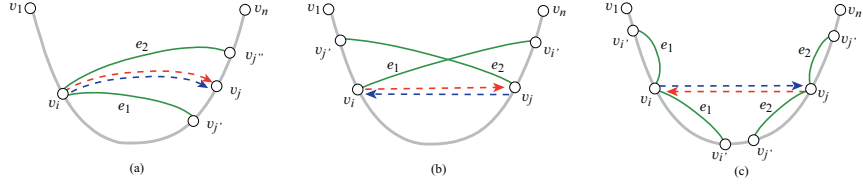} \end{center}
 \caption{Illustration of a gap  $\{v_i,v_j\}$ in an ordered graph $G=(V=\{v_1,v_2,\ldots,v_n\},E)$:
 (a) $e_1=v_i v_{j'}$ and $e_2=v_i v_{j''}$ such that $j'<j<j''$,
 (b)  $e_1=v_{i}v_{i'}$ and $e_2=v_{j}v_{j'}$ such that $j'<i$ and $j<i'$,
 (c)  $e_1=v_{i}v_{i'}$ and $e_2=v_{j}v_{j'}$ such that $i'<j$ and $i<j'$,
 where possibly $j'\leq i'$ or $i'<j'$.  } \label{fi:PCG_gap}
 \end{figure}

For an ordering $\sigma$ of $V$,   a non-adjacent vertex pair
$\{v_i,v_j\}$ with $i< j$ in $G$
is called a {\em gap} (with respect to edges $e_1,e_2\in E$)
if there are edges $e_1,e_2\in E$ that satisfy one of the following:\\
\hspace*{1mm} (g1)   $e_1=v_i v_{j'}$ and $e_2=v_i v_{j''}$ such that $j'<j<j''$
                    (or $e_1=v_{i'} v_{j}$ and \\
\hspace*{4mm} $e_2=v_{i''} v_{j}$   such that $i'<i<i''$), as illustrated  in Fig.~\ref{fi:PCG_gap}(a);   \\
\hspace*{1mm} (g2) $e_1=v_{i}v_{i'}$ and $e_2=v_{j}v_{j'}$ such that $j'<i$ and $j<i'$, as illustrated \\
\hspace*{4mm} in Fig.~\ref{fi:PCG_gap}(b); and \\
\hspace*{1mm} (g3)  $e_1=v_{i}v_{i'}$ and $e_2=v_{j}v_{j'}$ such that $i'<j$ and $i<j'$, as illustrated \\
\hspace*{4mm}  in Fig.~\ref{fi:PCG_gap}(c). \\
We call an ordering $\sigma$ of $V$ {\em gap-free} in $G$ if it has no gap.
Clearly the reversal of a gap-free ordering of $V$ is also gap-free.
We can test if a given ordering is gap-free or not
in $O(n^4)$ time by checking the conditions (a)-(c) for each non-adjacent vertex pair $\{v_i,v_j\}$ in $G$.

 Fig.~\ref{fi:PCG_example}(a) and (b)  illustrate the same graph $G_1$ with
 different orderings $\sigma_a=u_1,u_2,\ldots,u_8$ and $\sigma_b=v_1,v_2,\ldots,v_8$,
 where $\sigma_a$ is not gap-free while $\sigma_b$ is gap-free.

We have the following result, which implies that
a graph $G=(V,E)$ is a star-PCG if and only if
it admits a   gap-free ordering of $V$.

\begin{theorem}\label{thm:order}
For a graph $G=(V,E)$, let $\sigma$  be an ordering  of $V$.
Then there is  a PCR $(T_V,w,d_{\min},d_{\max})$ of $G$ such that
$w_1\leq w_2\leq\cdots\leq w_n$ if and only if
$\sigma$ is gap-free.
\end{theorem}


The necessity of this theorem is
relatively
easy to prove (see Lemma~\ref{le:necessity} in the Appendix).
Next we consider the sufficiency of Theorem~\ref{thm:order},
which is implied by the next lemma.

\begin{lemma}\label{le:compute_weight}
For a graph $G=(V,E)$, let 
  $\sigma=v_1,v_2,\ldots,v_n$
be an gap-free ordering  of $V$.
There is a PCR $(T_V,w,d_{\min},d_{\max})$ of $G$ such that
$w_1\leq w_2\leq\cdots\leq w_n$.
Such a set $\{w_1,w_2,\ldots,w_n, d_{\min},d_{\max}\}$ of
weights and bounds can be obtained in $O(n^3)$ time.
\end{lemma}

Note that when two vertices $u$ and $v$ are not adjacent in a PCG $G$,
there are two reasons: one is that the distance between them
 in the PCR $(T,w,d_{\min} , d_{\max})$
is smaller than $d_{\min}$, and the other is that the distance is larger than $d_{\max}$.
Before we try to assign some value to each $w_i$, we first detect this by
coloring edges in
the complete graph $K_V=(V,E\cup \overline{E})$  on the vertex set $V$
obtained from a graph $G$ by adding an edge between each non-adjacent vertex pair in $G$,
where $\overline{E}={ {V}\choose{2}}\setminus E$.

For a function $c: E\cup \overline{E}\to\{{\tt red}, {\tt green}, {\tt blue}\}$,
we call an edge $e$ with $c(e)={\tt red}$ (resp., {\tt green} and {\tt blue})
  a  red (resp., green and blue) edge,
  and let  $E_{\tt red}$ (resp., $E_{\tt green}$  and $E_{\tt blue}$) denote the sets
of red (resp., green   and blue) edges.
We denote by $N_{\tt red}(v)$ the set of neighbors
of a vertex $v$ via red edges.
We define $N_{\tt green}(v)$  and $N_{\tt blue}(v)$ analogously.

A {\em coloring} of $G=(V,E)$ is defined to be
  a function $c: E\cup \overline{E}\to\{{\tt red}, {\tt green}, {\tt blue}\}$
such that $E_{\tt green}=E$.
When an ordering $\sigma$ of $V$ is fixed,
we simply write $(i,j)\in {\tt red}$ (resp., $(i,j)\in {\tt green}$ and $(i,j)\in {\tt blue}$)
  if an edge $v_i v_j\in E\cup\overline{E}$ is a red (resp., green and blue) edge.
For $(G,\sigma)$ and a coloring $c$ of $G$,
 we wish to determine weights $w_i$, $i-1,2,\ldots,n$ and bounds $d_{\min}$ and $d_{\max}$
so that the next holds: \\
\hspace*{3mm} $w_1\leq w_2\leq \cdots\leq w_n$; \\
\hspace*{3mm} $ d_{\min}\leq w_i+w_j\leq d_{\max}$ for   $(i,j)\in {\tt red}$; \\
\hspace*{3mm} $w_i+w_j< d_{\min}$ for  $(i,j)\in {\tt green}$; and \\
\hspace*{3mm} $w_i+w_j> d_{\max}$ for $(i,j)\in {\tt blue}$.\\
To have such a set $\{w_1,\ldots,w_n, d_{\min},d_{\max}\}$ of
values for an ordering $\sigma$ and a coloring $c$ of  $G$,
the coloring $c$ must satisfy the following conditions: \\
  each $v_i\in V$ admits integers
  $a(i),b(i)\in [1,n ]$ such that
\[\mbox{  $N_{\tt red}(v_i)=\{v_j\mid 1\leq j\leq a(i)-1\}\setminus \{v_i\}$ and
  $N_{\tt blue}(v_i)=\{v_j\mid b(i)+1\leq j\leq n\}\setminus \{v_i\}$, }\]
  where  $a(i)=1$ if $N_{\tt red}(v_i)=\emptyset$;
     $b(i)=n$ if $N_{\tt blue}(v_i)=\emptyset$; and
     $N_{\tt green}(v_i)=V \setminus ( N_{\tt red}(v_i)\cup N_{\tt blue}(v_i)\cup \{v_i\})
     = \{v_j\mid a(i)\leq j\leq b(i)\}\setminus \{v_i\}\setminus \{v_i\}$, and $N_{\tt green}(v_i)=\emptyset$
     if $b(i)<a(i)$.
%
%
  Such a coloring $c$ of $G$ is called  {\em proper} to $(G,\sigma)$.

\begin{lemma}\label{le:proper_color}
For a graph $G=(V,E)$ and  a  gap-free ordering $\sigma$  of $V$,
 there is a coloring $c$ of $G$ that is proper to $(G,\sigma)$,
 which can be found in  in $O(n^2)$ time.
\end{lemma}

Define integers $i_{\tt red}$ and $i_{\tt blue}$ as follows.
 \[ i_{\tt red} = \left\{
 \begin{array}{cc}
 \mbox{the largest index $i$  such that  $i<a(i)$} &
     \mbox{ if $E_{\tt red}\neq\emptyset$},  \\
 0 & \mbox{ if $E_{\tt red}=\emptyset$},
 \end{array}
  \right. \]
 \[ i_{\tt blue} = \left\{
 \begin{array}{cc}
 \mbox{the smallest index $i$ such that  $b(i)<i$} &
    \mbox{ if $E_{\tt blue}\neq\emptyset$},  \\
 n+1 & \mbox{ if $E_{\tt blue}=\emptyset$}.
 \end{array}
  \right. \]
  In other words, $i_{\tt red}\neq 0$ is the largest $i$ with $( i, i+1)\in {\tt red}$, and $i_{\tt red}<n$,
  whereas
  $i_{\tt blue}\neq n+1$ is the smallest $i$ with $( i-1, i)\in {\tt blue}$, and $i_{\tt blue}>1$.
  Given a graph $G$, a gap-free ordering $\sigma=v_1,v_2,\ldots,v_n$ of $V$,
  and a coloring $c$ proper to $(G,\sigma)$,
  we can find
  the  set $\{a(i),b(i) \mid i=1,2,\ldots,n\}\cup \{ i_{\tt red}, i_{\tt blue}\}  $ of indices
  in $O(n^2)$ time.
  We also compute the set $M_G$ of all mirror pairs in $O(n^3)$ time.
Equipped with above results, we can prove the sufficiency of Theorem~\ref{thm:order}  by designing an $O(n)$-time
algorithm that assigns the right values to
weights $w_1,w_2,\ldots,w_n$ in $T_V$. The details can be found in Appendix~\ref{appd-ss}.

\section{Recognizing Star-PCGS}\label{sec:recognize}

 Based on Theorem~\ref{thm:order},
 we can test whether a  graph $G=(V,E)$ is a star-PCG or not
 by generating all $n!$ orderings of $V$.
 In this section, we show that testing whether a graph  has
a gap-free ordering of $V$  can be tested in
polynomial time.

\begin{theorem}\label{th:test_gap-free}
Whether a given graph  $G=(V,E)$ with $n$ vertices
  has a gap-free ordering of $V$  can be tested in
$O(n^6)$ time.
\end{theorem}


In a graph $G=(V,E)$, let
$E^{\tt t}$  denote the union of edge sets of all cycles of length 3 in $G$,
 $V^{\tt t}$ denote the set of end-vertices of edges in $E^{\tt t}$, and
$N_G^{\tt t}(v)$ denote the set of neighbors $u\in N_G(v)$ of a vertex $v\in V$
such that $uv\in E^{\tt t}$.

 \begin{figure}[htbp] \begin{center}
 \includegraphics[scale=0.14]{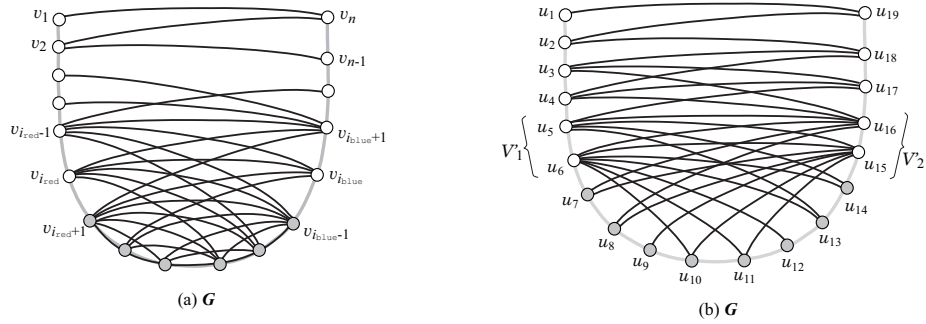} \end{center}
 \caption{
 (a) A disconnected graph $G=(V,E)$,
 (b) A connected non-bipartite graph $G=(V,E)$, where
 the edges between two vertices in $V^*=\{u_7,u_8,u_9,u_{10},u_{11},u_{12},u_{13},u_{14}\}$
 are not depicted.
   } \label{fi:PCG_star_structure2}
 \end{figure}

\begin{lemma}\label{le:structure1}
For a graph $G=(V,E)$ with
  a gap-free  ordering $\sigma= v_1,v_2,\ldots,v_n$  of  $V$ and
    a   coloring $c$  proper to $\sigma$,
    let   $V_1=\{v_i\mid 1\leq i\leq i_{\tt red}\}$,
     $V_2=\{v_i\mid i_{\tt blue}\leq i\leq n\}$, and
     $V^*=\{v_i \mid i_{\tt red}-1 \leq i\leq  i_{\tt blue}+1\}$.
Then
\begin{enumerate}
\item[{\rm (i)}]
If two edges $v_i v_j$ and $v_{i'}  v_{j'}$
 with $i<j$ and $i'<j'$   cross $($i.e.,   $i<i'<j<j'$ or $i'<i<j'<j)$,
 then they belong to the same component  of $G$;
\item[{\rm (ii)}] It holds
 $i_{\tt red}+1 \leq i_{\tt blue}-1$.
 The graph $G[V^*]$ is a complete graph,
 and $G-V^*$ is a bipartite graph between vertex sets  $V_1$ and $V_2$;
\item[{\rm (iii)}]
Every two vertices $v_i,v_j\in V_1\cap N_G(V^*)$ with $i<j$
satisfy  $v_{i_{\tt blue}-1}\in N_G(v_i)\cap V^*\subseteq N_G(v_j)\cap V^*
\subseteq V^*\setminus \{v_{i_{\tt red}+1}\}$; and \\
Every two vertices $v_i,v_j\in V_2\cap N_G(V^*)$ with $i<j$
satisfy   $v_{i_{\tt red}+1}\in N_G(v_j)\cap V^*\subseteq N_G(v_i)\cap V^*
\subseteq V^*\setminus \{v_{i_{\tt blue}-1}\}$.
\end{enumerate}
\end{lemma}

We call the complete graph $G[V^*]$ in Lemma~\ref{le:structure1}(ii) the {\em core} of $G$.
Based on the next lemma,
we can treat each component of a disconnected graph $G$ separately
to test whether $G$ is a star-PCG or not.

\begin{lemma}\label{le:structure11}
Let $G=(V,E)$ be a   graph with at least two components.
\begin{enumerate}
\item[{\rm (i)}]
If $G$ admits a gap-free  ordering  of  $V$, then
each component  of $G$ admits a gap-free  ordering of its vertex set,
and there is at most one non-bipartite component in $G$; and
\item[{\rm (ii)}]
Let $G'=(V'_1,V'_2,E')$ be a bipartite component of $G$, and $G''=G-V(G')$.
Assume that $G'$
admits  a gap-free  ordering $v'_1,v'_2,\ldots,v'_p$  of  $V'_1\cup V'_2$
and $G''$ admits  a gap-free  ordering $v_1,v_2,\ldots,v_q$  of  $V_2$.
Then there is an index $k$ such that
 $\{\{v'_1,v'_2,\ldots,v'_k\},\{v'_{k+1}, v'_{k+2}, \ldots, v'_p\}\}=\{V'_1, V'_2\}$.
Moreover, the ordering $v'_1,v'_2,\ldots,v'_k,v_1,v_2,\ldots,v_q,v'_{k+1}, v'_{k+2}, \ldots, v'_p$ of $V$
is gap-free to $G$.
\end{enumerate}
\end{lemma}
\noindent {\bf Proof.}
(i) Let  $G$ admit  a gap-free  ordering  of  $V$.
Any induced subgraph $G$ such as a component of $G$
 is a star-PCG, and a gap-free  ordering of its vertex set
by Theorem~\ref{thm:order}.
By Lemma~\ref{le:structure1}(i), at most one component $H$ containing
a complete graph with at least three
vertices can be non-bipartite, and the remaining graph $G-V(H)$ must
 be a collection of bipartite graphs.

(ii) Immediate from the definition of gap-free orderings.
\qed \bigskip

We first consider the problem of
testing if a given connected bipartite graph is a star-PCG  or not.
We   reduce this
to the problem of finding contiguous ordering to a family of sets.
For a  bipartite graph $G=(V_1, V_2, E)$,
define  $\mathcal{S}_i$ to be the family $\{N_G(v)\mid v\in V_j\}$
 for the $j\in \{1,2\}-\{i,j\}$, where even if
there are distinct vertices $u,v\in V_j$ with  $N_G(u)=N_G(v)$,
  $\mathcal{S}_i$ contains exactly one set   $S=N_G(u)=N_G(v)$.

 For the example of a connected bipartite graph $G_1=(V_1,V_2,E)$ in Fig.~\ref{fi:PCG_example}(a),
  we have
 $\mathcal{S}_1=\{
 \{v_3,v_4\},
 \{v_1,v_2,v_3,v_4\}
  \}$,
  and
 $\mathcal{S}_2=\{
 \{v_5,v_6\},
 \{v_5,v_6,v_7,v_8\}
 \}$.

\begin{lemma}\label{le:bipartite_case}
Let $G=(V_1,V_2,E)$ be a connected  bipartite  graph with $|E|\geq 1$.
Then  family $\mathcal{S}_i$ is separator-free for each $i=1,2$,
and $G$ has a gap-free ordering of $V$
 if and only if for each $i=1,2$, family $\mathcal{S}_i$ admits a contiguous ordering $\sigma_i$ of $V_i$.
 For any contiguous ordering $\sigma_i$ of $V_i$, $i=1,2$,
 one of orderings  $(\sigma_1,\sigma_2)$ and  $(\sigma_1, \overline{\sigma_2})$
  of $V$  is  a gap-free ordering to $G$.
\end{lemma}

 Note that
 $|\mathcal{S}_1|+|\mathcal{S}_2|+|V(\mathcal{S}_1)|+|V(\mathcal{S}_2)|=O(n)$.
 By Theorem~\ref{th:contiguous},
 a contiguous ordering of $V(\mathcal{S}_i)$ for each $i=1,2$ can be computed
  in $O(|V(\mathcal{S}_i)||\mathcal{S}_i|^2)=O(n^3)$ time.

 Fig.~\ref{fi:PCG_example}(a) illustrates  an ordering
 $\sigma_a=v_1,v_2,v_3,v_4,v_8,v_7,v_6,v_5$ of $V(G_1)$ of a connected bipartite graph $G_1=(V_1,V_2,E)$,
 where $\sigma_a$ consists of a contiguous ordering $\sigma_1=v_1,v_2,v_3,v_4$ of $V_1$
 and a contiguous ordering $\sigma_2=v_8,v_7,v_6,v_5$ of $V_2$.
 Although $\sigma_a$ is not gap-free in $G$,
 the other ordering $\sigma_b$ of $V(G_1)$ that consists of $\sigma_1$ and
  the reversal of $\sigma_2$
 is gap-free, as illustrated in Fig.~\ref{fi:PCG_example}(b).


Finally we consider the case where a given graph $G$ is a
connected and non-bipartite graph.
 Fig.~\ref{fi:PCG_example}(d) illustrates a
connected and non-bipartite star-PCG whose maximum clique is not unique.


\begin{lemma}\label{le:structure2}
For a connected non-bipartite  graph  $G=(V,E)$  with $V^{\tt t}\neq \emptyset$,
and let $v^*_1,v^*_2$ be two adjacent vertices in $V^{\tt t}$.
Let $V^*=\{v^*_1,v^*_2\}\cup(N_G(v^*_1)\cap N_G(v^*_2))$,
$V'_1=N_G(v^*_2)\setminus V^*$, and
$V'_2=N_G(v^*_1)\setminus V^*$.
Assume that $G$ has a gap-free ordering $\sigma$ of $V$ and a proper coloring $c$ to $\sigma$
 such that
$v^*_1=v_{i_{\tt red}+1}$, $v^*_2=v_{i_{\tt blue}-1}$.
Then:
\begin{enumerate}
\item[{\rm (i)}]
A maximal clique $K_{v^*_1,v^*_2}$ of $G$ that contains edge $v^*_1,v^*_2$
  is uniquely given as
  $G[V^*]$.
The graph $G[V^*]$ is the core of  the ordering $\sigma$,
 and $G-V^*$ is a bipartite graph $(V_1,V_2,E')$; and
\item[{\rm (ii)}]
Let $\mathcal{S}_i$ denote the family $\{N_G(v)\mid v\in V_j\}$ for $\{i,j\}=\{1,2\}$,
 and  $\mathcal{S}=\mathcal{S}_1\cup \mathcal{S}_2\cup\{ V^* \}$.
Then
$\mathcal{S}$ is a separator-free family
that admits a contiguous ordering $\sigma$ of $V$,
  and any  contiguous ordering $\sigma$ of $V$ is  a gap-free ordering to $G$.
\end{enumerate}
\end{lemma}

 For example,
 when we choose  vertices
 $v^*_1=u_7$ and  $v^*_2=u_{14}$ in
 the connected non-bipartite graph $G=(V,E)$ in Fig.~\ref{fi:PCG_star_structure2}(b),
  we have
 $V^*=\{u_7,u_8,u_9,u_{10},u_{11},$ $u_{12}, u_{13},u_{14}\}$,
 $\mathcal{S}_1=\{
 \{u_1,u_2\}, $ $
 \{u_2,u_3,u_4\},
 \{u_3,u_4,u_5\},$ $
 \{u_3,u_4,u_5,u_6,u_7,u_8\},$ $
 \{u_5,u_6,u_7,u_8,u_9,u_{10},u_{11}\}
  \}$,
  and
 $\mathcal{S}_2=\{
 \{u_{19}\},$ $
 \{u_{18},u_{19}\},
 \{u_{16},u_{17},u_{18}\},$ $
 \{u_{13},$ $u_{14},u_{15},u_{16},u_{17}\},$ $
 \{u_{10},u_{11},u_{12},u_{13},u_{14},u_{15},u_{16}\}
 \}$.

 For a fixed $V^*$ in Lemma~\ref{le:structure2}, we can test
 whether the separator-free family $\mathcal{S}$
 in Lemma~\ref{le:structure2}(ii) is constructed from $V^*$
 in  $O(|V(\mathcal{S})| |\mathcal{S}|^2)= O(n^3)$ time by Theorem~\ref{th:contiguous},
 since $|\mathcal{S}|+|V(\mathcal{S})|=O(n)$ holds.
 It takes $O(n^4)$ time to check a given ordering is gap-free or not.
 To find the right choice of a vertex pair
$v^*_1=v_{i_{\tt red}+1}$ and $v^*_2=v_{i_{\tt blue}-1}$ of some gap-free ordering $\sigma$ of $V$,
 we need to try $O(n^2)$ combinations of vertices to construct $V^*$ according to the lemma.
 Then we can find  a gap-free ordering of a given graph, if one exists in $O(n^6)$ time,
 proving Theorem~\ref{th:test_gap-free}.

By Theorems~\ref{thm:order} and \ref{th:test_gap-free},
we conclude that whether a given graph with $n$ vertices  is a star-PCG or not can be
tested in $O(n^6)$ time.

\section{Concluding Remarks}\label{sec:conclude}
Pairwise compatibility graphs were initially introduced from the context of phylogenetics in computational biology
and later became an interesting graph class in graph theory.
PCG recognition is a hard task and we are still far from a complete characterization of PCG.
Significant progresses toward PCG recognition would be interesting from a graph theory perspective and also be helpful in designing
sampling algorithms for phylogenetic trees. In this paper, we give the first polynomial-time algorithm to recognize star-PCGs.
Although stars are trees of a simple topology, it is not an easy task to recognize star-PCGs.
For further study, it is an interesting topic to study the characterization of PCGs with witness trees of other particular topologies.

\newpage

\appendix
\centerline{\bf\large Appendix}

\section{Proof of Lemma~\ref{lem_ini}}

\textbf{Lemma~\ref{lem_ini}}
\emph{Every PCG admits a PCR $(T,w,d_{\min}, d_{\max})$ such that
  $0<d_{\min}< d_{\max}$ and 
  for all edges $e\in E(T)$. }
  \medskip
  
\noindent {\bf Proof.}
Since  the case where a given PCG $G$ has at most two vertices is trivial,
we assume that $G$ has at least three vertices.
Let $(T,w',d'_{\min}, d'_{\max})$ be a PCR of $G$,
where  each path between two leaves in $T$ contains exactly two leaf edges  
since  $|L(T)|=|V(G)|\geq 3$.
We increase some values of $w'(e)$, $e\in E(T)$ and $d'_{\min}$ and $d'_{\max}$ 
so that the resulting tuple  satisfies the lemma. 
For two positive reals $\delta$ and 
   $\varepsilon<\min_{uv\not\in E}\mathrm{d}_{T,w'}(u,v) -d'_{\max}$,
  let $w(e):=w'(e)+\delta$ for each leaf edge $e\in E(T)$, 
   $w(e):=w'(e)$ for all non-leaf edges $e\in E(T)$, 
   $d_{\min}:=d'_{\min}+2\delta$, and $d_{\max}:=d'_{\max}+2\delta+ \varepsilon$.
 We observe that  $(T,w, d_{\min},d_{\max})$ is a PCR of $G$
satisfying the lemma because  
    each path between two leaves in $T$ contains exactly two leaf edges 
 \qed

\section{Proof of Theorem~\ref{th:consecutive}}\label{proofTh1}

First we prove the time complexity of the theorem. 
A graph $H$ with $n\geq 1$ vertices is called an interval graph if
each vertex $v\in V(H)$ can represented by an ordered pair $(a_v,b_v)$ of 
reals $a_v\leq b_v$  so that 
two vertices $u,v\in V(H)$ are adjacent if and only if there is a real $c$
such that two intervals $a_u\leq c\leq b_u$ and $a_v\leq c\leq b_v$.
It is know that testing if a given graph $H$ is an interval graph
and finding such a representation $(a_v,b_v)$, $v\in V(H)$, if one exists
can be done in $O(|V(H)|+|E(H)|)$ time~\cite{bl}.
Given a  family $\mathcal{S}\subseteq 2^V$, 
we see that $\mathcal{S}$ admits a consecutive ordering
if and only if the auxiliary graph $H_{\mathcal{S}}$ for $\mathcal{S}$
is an interval graph by the definition of $H_{\mathcal{S}}$.
The time to construct $H_{\mathcal{S}}$ from  $\mathcal{S}$
is  $O(nm^2)$ since  we can check in $O(n)$  time
 whether two sets $S,S'\in \mathcal{S}$
 intersect, i.e., vertices $S,S'\in V(H)$ are adjacent in $H$.
Clearly $|V(H)|=|\mathcal{S}|=m$ and $|E(H)|\leq |\mathcal{S}|^2=O(m^2)$.
Then testing if $H_{\mathcal{S}}$ is an interval graph and
finding  a representation $(a_v,b_v)$, $v\in V(H)$ can be done in 
$O(n+m^2)$ time.
In total, it takes $O(nm^2)$ time to 
 find a consecutive ordering for $\mathcal{S}$, if one exits.  
  
Next we prove that  
  a consecutive ordering  of $V$ to a cut-free family $\mathcal{S}$ is unique up to reversal.
Assume that $\mathcal{S}$ admits another consecutive ordering $\tau\not\in\{\sigma,\overline{\sigma}\}$
to derive a contradiction that $\mathcal{S}$ would have a cut. 
Since $\tau\not\in\{\sigma,\overline{\sigma}\}$, 
some two elements $a,b\in V$ appear consecutively in $\sigma$, 
but another element $c\in V\setminus\{a,b\}$ appears 
in the subsequence $\tau_{ab}$ of $\tau$ between $a$ and $b$.
Hence 
$a=v_k$, $b=v_{k+1}$ and $c=v_p$ for $\sigma=v_1, v_2,\ldots, v_n$,
where $k+1<p$ is assumed without loss of generality and
  $c=v_p$ is chosen from $\tau_{ab}$ so that the index $p$ is maximized.
We claim that the set $C=\{v_{k+1}, v_{k+2}, \dots, v_{p}\}$ is a cut to $\mathcal{S}$. 
To derive a contradiction, $C$ intersects  a set $S\in \mathcal{S}$, where 
  $S=\{v_s, v_{s+1},\ldots, v_{t}\}$ 
for 
$s\leq k<k+1\leq t<p$ or $ k+1<s\leq p<t$ by the consecutiveness of $S$ in $\sigma$.
In the former, $S$ with $a,b\in S$   contains  $\tau_{ab}$, implying $c=v_p\in S$, a contradiction.
In the latter, $S$ is contained in $\tau_{ab}$, since $a,b\not\in S$ and $c\in S$, 
implying that   $v_t$ with $p<t$ belongs to  $\tau_{ab}$, 
 a contradiction to  the choice of  $c= v_p$. 
\bigskip

\section{Proof of Theorem~\ref{th:contiguous}}\label{appd2}

To prove Theorem~\ref{th:contiguous}, we show that an instance $(V,\mathcal{S})$ of the problem of finding
a  contiguous ordering of $V$ to $\mathcal{S}$ can be modified to
an instance $(V,\mathcal{S}')$ of the problem of finding
a  consecutive ordering of $V$ to $\mathcal{S}'$
by introducing some new subsets  of $V$.

A set $S\in \mathcal{S}$ is called {\em minimal} (resp., {\em maximal})
if $\mathcal{S}$ contains no proper subset (resp., superset) of $S$. 
Define $\mathcal{S}_{\min}$ (resp., $\mathcal{S}_{\max}$)
to be the family of minimal (resp.,   maximal) sets in $\mathcal{S}$.
 
 Theorem~\ref{th:contiguous} follows from (iv) and (v) of the next lemma. 

\begin{lemma}\label{le:conversion}
For a set $V$ of  $n\geq 1$ elements, 
let  $\mathcal{S}\subseteq 2^V\setminus \{\emptyset\}$ be a
 family   with $m\geq 1$ sets.
 \begin{enumerate}
 \item[{\rm (i)}] 
If some set $B\in\mathcal{S}_{\max}$ contains at least three sets from $\mathcal{S}_{\min}$, 
 then $\mathcal{S}$ cannot have a contiguous ordering of $V$;
 \item[{\rm (ii)}]
 Assume that $\mathcal{S}$ is separator-free.
 Let $X$ be a maximal set of elements any two of which are equivalent.
 Then the elements in $X$ appear consecutively in any contiguous ordering of $V$ to $\mathcal{S}$; 
 \item[{\rm (iii)}]
 Assume that $\mathcal{S}$ is  simple and separator-free.
 Let $\mathcal{S}'$  
 denote the
  family obtained from $\mathcal{S}$ by adding
  a new set $S_{A,B}=B\setminus A$
for each pair of   sets $A\in  \mathcal{S}_{\min}$ and $B\in\mathcal{S}_{\max}$
such that  
$A\subsetneq B$ 
  and
    $B\setminus A\not\in \mathcal{S}$.
    Then $\mathcal{S}'$ is cut-free and 
      any  consecutive ordering of $V$ to $\mathcal{S}'$ is 
    a contiguous ordering of $V$ to $\mathcal{S}$; and 
 \item[{\rm (iv)}]
 Assume that $\mathcal{S}$ is separator-free.
 Then a contiguous ordering of $V$ to $\mathcal{S}$ can be found in 
 $O(nm^2)$  time, if one exists.
 Moreover 
 all elements in each set $X\in \mathcal{X}_{\mathcal{S}}$ appear
 consecutively in any contiguous ordering of $V$ to $\mathcal{S}$,
 and a  contiguous ordering of $V$ to $\mathcal{S}$ is unique up to reversal 
 of the entire ordering and
 arbitrariness of orderings of elements in each set $X\in \mathcal{X}_{\mathcal{S}}$; and
 \item[{\rm (v)}]
 A contiguous ordering of $V$ to $\mathcal{S}$ can be found in 
 $O(nm^2)$  time, if one exists.
 \end{enumerate}
\end{lemma}
\noindent {\bf Proof.}
(i) Let  three sets  $A_i\in  \mathcal{S}_{\min}$, $i=1,2,3$ be contained in some set $B\in\mathcal{S}_{\max}$.
Note that  each $A_i$ is a proper subset of $B$, and no set $A_i$ is contained in any other set $A_j$ with $j\neq i$.
Hence in any  contiguous ordering of $V$ to $\mathcal{S}$,
at least one of the three sets $A_i$, $i=1,2,3$ cannot share
the first element in $B$ or the last element in $B$.
This means that  $\mathcal{S}$ cannot have a contiguous ordering of $V$.

(ii) To derive a contradiction, assume that there is a contiguous ordering
$u_1,u_2,\ldots,u_n$ of $V$ to $\mathcal{S}$
wherein the indices of elements $X$ are not consecutive,
i.e., there are elements $u_i,u_j,u_k\in V$ with $i<j<k$
and a set $S\in \mathcal{S}$ such that
$\{u_i,u_j,u_k\}\cap S=\{u_j\}$.
Since $\mathcal{S}$  is separator-free, there is a set $S'\in \mathcal{S}$ that intersects
$S$ or contains $S$.
Since any two elements in $X$ are equivalent, it holds that either $X\subseteq S'$ or $X\cap S'=\emptyset$.
If $S'$ intersects $S$, then $S'\setminus S\neq\emptyset$ means that
$S'\supseteq X$ and $S'$ would contain $S$ too since the indices of elements in $S'$ are
consecutive in the ordering.
Hence $S'$ always contains $S$, and it also contains $X$,
where  $S'\setminus S\supseteq X$.
Now $S$ does not contain the first element or the last element of $S'$ in the ordering.
This contradicts that the ordering is contiguous   to $\mathcal{S}$.

(iii)  For any two  sets $A\in  \mathcal{S}_{\min}$ and $B\in\mathcal{S}_{\max}$
such that $A\subseteq B$ and  $B\setminus A\neq\emptyset$,
the set
    $B\setminus A$ must consist of elements with consecutive indices
    in a contiguous ordering of $V$ to $\mathcal{S}$.
Hence after adding such a set $B\setminus A$ to $\mathcal{S}$,
the  contiguous ordering of $V$ to $\mathcal{S}$ is a  consecutive ordering of $V$ to $\mathcal{S}'$.
We show that any consecutive ordering of $V$ to $\mathcal{S}'$ is a contiguous  ordering of $V$ to $\mathcal{S}'$.
Assume that, for a  consecutive ordering $u_1,u_2,\ldots,u_n$
of $V$ to $\mathcal{S}'$, there are sets $X,Y\in \mathcal{S'}$ 
such that $X\subseteq Y=\{u_i,u_{i+1},\ldots,u_{i+|Y|-1}\}$
but $X$ does not contain any of $u_i$ and $u_{i+|Y|-1}$.
Let  $A_X\in  \mathcal{S}_{\min}$ be a set such that $A_X\subseteq X$
and let $B_Y\in\mathcal{S}_{\max}$ be a set such that $Y\subseteq B_Y$.
Note that   $\mathcal{S'}$ contains $B_Y\setminus A_X$,
where  $u_i, u_{i+|Y|-1}\in B_Y\setminus A_X$.
However $B_Y\setminus A_X$ does not consist of elements with consecutive indices
    in a the consecutive ordering $u_1,u_2,\ldots,u_n$, a contradiction.
    Hence  any consecutive ordering of $V$ to $\mathcal{S}'$
     is a contiguous  ordering of $V$ to $\mathcal{S}'$.

We next prove that $\mathcal{S}'$ is cut-free.
Let $C$ be a non-trivial subset of $V$, and assume that no set  $\mathcal{S}$ intersects $C$.
Since $|C|\geq 2$ and $\mathcal{S}$ is simple,
there  is a set $S\in \mathcal{S}$ such that $S\cap C\neq\emptyset \neq C\setminus S$.
If $S\setminus C\neq\emptyset$, then $S$ would intersect  $C$.
Hence $S\setminus C$.
If each set $S'\in \mathcal{S}$ with $S'\cap (V\setminus C)\neq\emptyset$ is
disjoint with $C$, then this would contradict that $\mathcal{S}$
is separator-free.
Hence there is a set $S'\in \mathcal{S}$ with $S'\cap (V\setminus C)\neq\emptyset$
and $S'\supseteq C$.
After the above procedure,  the resulting family $\mathcal{S}'$ contains
a set $S''$ such that $S\setminus S''\neq\emptyset$ and $S''\subseteq S'\setminus S$,
where $C\cap S''\supseteq C\setminus S\neq\emptyset$,
$C\setminus S''\supseteq S\setminus S''\neq\emptyset$, and
$S''\setminus C\supseteq S'\cap (V\setminus C)\neq\emptyset$.
Therefore $S''$ intersects $C$.
This proves that   $\mathcal{S}'$ is cut-free.

(iv) 
  Let  $\mathcal{S}_0\subseteq 2^V$ be a given separator-free family
  with $m\geq 1$ subsets. 
First we compute the family $\mathcal{X}_{\mathcal{S}_0}$ of  all maximal subsets. 
This takes  $O(n m^2)$ time. 
By (ii) of this lemma,
all elements in each $X\in \mathcal{X}_{\mathcal{S}_0}$ appear consecutively
in any contiguous ordering of $V$.
Next we construct a simple family $\mathcal{S}$ from $\mathcal{S}_0$
as follows. 
 For each set $X\in \mathcal{X}_{\mathcal{S}_0}$,
 we choose one element $v_X$ from $X$ and replace $X$ with $X\setminus\{u_X\}$
 in the family.
 Note that  the resulting family $\mathcal{S}$ is  simple and
separator-free and $|\mathcal{S}|\leq |\mathcal{S}_0|$. 
If $\mathcal{S}_0$ admits a  contiguous ordering of $V$ then so does $\mathcal{S}$. 
For the family $\mathcal{S}$, 
we then compute  the families $\mathcal{S}_{\min}$ and $\mathcal{S}_{\max}$
 and  construct 
  a bipartite graph $\mathcal{B}=( \mathcal{S}_{\min},\mathcal{S}_{\max}, E^*)$
 such that
 for each pair of sets  $A\in \mathcal{S}_{\min}$,  $B\in \mathcal{S}_{\max}$,
  $AB\in E^*$ if and only if $A\subseteq B$.
  This also can be done in  $O(n m^2)$ time. 
Now testing whether there is a set $B\in \mathcal{S}_{\max}$ satisfying
the condition (i) of this lemma
 can be done in  $O(m^2n)$ time, because  a set  $B\in \mathcal{S}_{\max}$ contains
three sets in $\mathcal{S}_{\min}$ if and only if 
  the degree 
 of  vertex $B\in \mathcal{S}_{\max}$ in $\mathcal{B}$ 
 is at least 3. 
 When there exists such a set  $B\in \mathcal{S}_{\max}$,
 we conclude that $\mathcal{S}_0$ does not admits any contiguous ordering of $V$.
Assume that no  set $B\in \mathcal{S}_{\max}$ satisfies 
the condition (i) of this lemma,
and
construct from $ \mathcal{S}$ the cut-free family  $\mathcal{S}'$
 in (iii) of this lemma,
where we see that
$|\mathcal{S}'|\leq |\mathcal{S}|+  2|\mathcal{S}_{\max}|\leq 3m$. 
By Theorem~\ref{th:consecutive} applied to $\mathcal{S}'$,
we can find a consecutive ordering $\sigma$ of $V(\mathcal{S}')$ to
the cut-free family $\mathcal{S}'$ in $O(m^2n)$ time, if one exists,
where  $\sigma$ is unique up to reversal.
By  (iii) of this lemma,,
the ordering  $\sigma$  is contiguous to $\mathcal{S}$.
All contiguous orderings of $V$ to the input separator-free family $\mathcal{S}_0$
can be obtained by replacing the representative element $v_X$ in $\sigma$ 
 for each set $X\in \mathcal{X}_{\mathcal{S}_0}$ with an arbitrary ordering of $X$.
 Therefore a contiguous ordering $\sigma^*$ of $V$ to $\mathcal{S}_0$ 
 can be found in $O(nm^2)$ time, if one exists, and  $\sigma^*$
 is unique up to reversal and arbitrariness of 
 orderings of elements in each set $X\in \mathcal{X}_{\mathcal{S}_0}$. 
 
 (v) 
Given a family $\mathcal{S}\subseteq 2^V$, 
we can test in $O(nm)$ time if a set $X\in \mathcal{S}$ is a 
 cut of $\mathcal{S}$, i.e.,
  $X$ does not intersect any other set $S\in \mathcal{S}\setminus\{X\}$.
  Then the family $\mathcal{C}_{\mathcal{S}}$ of all  cuts  of $\mathcal{S}$ can be found in $O(nm^2)$ time,
  where $|\mathcal{C}_{\mathcal{S}}|\leq 2n$ since $\mathcal{C}_{\mathcal{S}}$ is a laminar. 
Note that any inclusion-wise maximal set in $\mathcal{C}_{\mathcal{S}}$ is a separator of $\mathcal{S}$.
For each cut $X\in  \mathcal{C}_{\mathcal{S}}$, let $\mathcal{C}(X)$ denote
the  family of sets $Y\in \mathcal{C}_{\mathcal{S}}$ such that
$Y\subsetneq X$ and $Y$ is maximal, i.e., no other set $S\in \mathcal{C}_{\mathcal{S}}$
satisfies $Y\subsetneq S\subsetneq X$.
For each set $X\in \mathcal{C}_{\mathcal{S}}$,
let 
$\mathcal{S}[X]$ denote the family of sets $S \in\mathcal{S}$
with $S\subseteq X$, and 
$\mathcal{S}\langle X\rangle$ denote the family obtained
from  
$\mathcal{S}[X]$ by contracting each set $Y\in \mathcal{C}(X)$
  into a single element $v_Y$,
ignoring all sets $S\in \mathcal{S}$ with $S\subseteq Y$,
where  
$|V(\mathcal{S}\langle X\rangle)|=|X|-\sum_{Y\in \mathcal{C}(X)}(|Y|-1)$
and 
$|\mathcal{S}\langle X\rangle|=|\mathcal{S}[X]| 
-\sum_{T\in \mathcal{C}(X)}|\mathcal{S}[Y]|$. 
We easily see that 
$\sum_{X\in \mathcal{C}_{\mathcal{S}} }|V(\mathcal{S}\langle X\rangle)|
 \leq n+|\mathcal{C}_{\mathcal{S}}|\leq 3n$, and 
$\sum_{X\in \mathcal{C}_{\mathcal{S}} }|\mathcal{S}\langle X\rangle|\leq m$.
%
Observe that, for each set $X\in\mathcal{C}_{\mathcal{S}}$, the family $\mathcal{S}\langle X\rangle$ 
 is separator-free, and
 $\mathcal{S}[X]$ admits a contiguous ordering of $V$ if and only if
 $\mathcal{S}\langle X\rangle$ admits a contiguous ordering of $V(\mathcal{S}\langle X\rangle)$
 and 
$\mathcal{S}[Y]$ for each set $Y\in \mathcal{C}(X)$
 admits a contiguous ordering of $V(\mathcal{S}[Y])$.
To construct a contiguous ordering of $V$ to $\mathcal{S}$,
we choose each set  $X\in\mathcal{C}_{\mathcal{S}}$ in a non-decreasing order of size $|X|$, where  contiguous orderings $\sigma_{[Y]}$ for families $\mathcal{S}[Y]$ with sets $Y\in \mathcal{C}(X)$
are available by induction.
We then  find 
  a contiguous ordering $\sigma_{\langle X\rangle}$ for family  $\mathcal{S}\langle X\rangle$ 
    in    $O(|V(\mathcal{S}\langle X\rangle)||\mathcal{S}\langle X\rangle|^2)$ time, if one exists 
   by (iv) of this lemma,
   and  
construct a contiguous ordering  $\sigma_{[X]}$ for  $\mathcal{S}[X]$
by replacing the element $v_Y$ in $\sigma_{\langle X\rangle}$ with ordering  $\sigma_{[Y]}$ 
for each set $Y\in \mathcal{C}(X)$  in $O(n)$ time.
 Since $\sum_{X\in \mathcal{C}_{\mathcal{S}} }|V(\mathcal{S}\langle X\rangle)||\mathcal{S}\langle X\rangle|^2 =O(nm^2)$,
 we can find a contiguous ordering of $V$ to $\mathcal{S}$ in $O(nm^2)$ time, if one exists. 
\qed \bigskip

\section{The Necessity of Theorem~\ref{thm:order}}

The necessity of Theorem~\ref{thm:order} is given by the following lemma.

\begin{lemma}\label{le:necessity}
For a graph $G=(V,E)$, let $\sigma=v_1,v_2,\ldots,v_n$  be an ordering  of $V$.
If $\sigma$ is not gap-free, then
 there is no PCR $(T_V,w,d_{\min},d_{\max})$ of $G$ such that
$w_1\leq w_2\leq\cdots\leq w_n$.
\end{lemma}

\noindent {\bf Proof.}
Let  $(G,\sigma)$
  admit a PCR $(T_V,w,d_{\min},d_{\max})$ with $w_1\leq w_2\leq\cdots\leq w_n$.
To derive a contradiction, assume that $\sigma$ has a gap $\{v_i,v_j\}$ with $i<j$.
If it satisfies the condition (g1) with respect to edges $e_1=v_i v_{j'}$ and $e_2=v_i v_{j''}$ such that $j'<j<j''$
(or $e_1=v_{i'} v_{j}$ and $e_2=v_{i''} v_{j}$ such that $i'<i<i''$),
then  $d_{\min}\leq w_i+w_{j'}\leq w_i+w_j\leq w_i+w_{j''}\leq d_{\max}$
(or $d_{\min}\leq w_{i'}+w_j\leq w_i+w_j\leq w_{i''}+ w_j\leq d_{\max}$),
which implies $v_iv_j\in E$, i.e., $v_i$ and $v_j$ must be adjacent in $G$, a contradiction.
If the gap satisfies the condition (g2) with respect to edges $e_1=v_{i}v_{i'}$ and
 $e_2=v_{j}v_{j'}$ such that $j'<i$ and $j<i'$,
then    $d_{\min}\leq w_{j'}+w_j\leq w_i+w_j\leq w_i +w_{i'}\leq d_{\max}$,
again implying that  $v_i$ and $v_j$ must be adjacent, a contradiction.
Analogously with the case where the gap satisfies the condition (g3)
with respect to edges $e_1=v_{i}v_{i'}$ and $e_2=v_{j}v_{j'}$ such that $i'<j$ and $i<j'$, where
  $d_{\min}\leq w_i +w_{i'}\leq w_i+w_j\leq  w_{j'}+w_j \leq d_{\max}$
would imply that  $v_i$ and $v_j$ are adjacent in $G$.
 \qed

\section{Proof of Lemma~\ref{le:compute_weight}: The Sufficiency of Theorem~\ref{thm:order}}\label{appd-ss}

For the sufficiency of Theorem~\ref{thm:order},
we prove Lemma~\ref{le:compute_weight}
by designing an $O(n)$-time algorithm that assigns the right  values to
weights $w_1,w_2,\ldots,w_n$ in $T_V$.

We start with proving Lemma~\ref{le:proper_color}.
\medskip


\noindent
\textbf{Lemma~\ref{le:proper_color}}
\emph{For a graph $G=(V,E)$, let $\sigma=v_1,v_2,\ldots,v_n$  be an ordering  of $V$.
For a graph $G=(V,E)$ and  a  gap-free ordering $\sigma$  of $V$,
 there is a coloring $c$ of $G$ that is proper to $(G,\sigma)$,
 which can be found in  in $O(n^2)$ time.}
\medskip

\noindent {\bf Proof.}
 First we assign color green all edges in $E$.
 Since  $(G,\sigma)$ has no gap in the condition (g1),
 the neighbor set $N_G(v_i)$ of each vertex $v_i\in V$
 is given by a set  $\{v_{a_i}, v_{a_i+1},\ldots, v_{b_i}\}\setminus\{v_i\}$
 of vertices  with consecutive indices.
 Next we assign color red to all edges $v_j v_i\in \overline{E}$ with $j<a_i$
 and color blue to all edges $v_iv_k\in \overline{E}$ with $b_i<k$.
 It suffices to show that no edge $v_iv_j\in \overline{E}$ with $i<j$ is assigned two colors at the same time,
in such a way that either \\
\hspace*{1mm}  (i)  color red from $v_i$ and color blue from $v_j$; or \\
\hspace*{1mm}  (ii) color blue from $v_i$ and color red from $v_j$.\\
When (i) (resp., (ii)) occurs,  there are edges
 $v_iv_{i'}\in E$ and $v_{j'}v_{j}\in E$ such that $j'<i<j<i'$
 (resp., $i'<j$ and $i<j'$),
 which means that $\{v_i,v_j\}$ would be a gap in the condition (g2) (resp., (g3)), a contradiction.
 Hence the above procedure constructs a coloring $c$  proper  to  $(G,\sigma)$.
 We easily see that the  procedure can be implemented to run in $O(n^2)$ time.
 \qed
\medskip

By the lemma, we consider the case where 
a given graph $G=(V,E)$ with an ordering $\sigma$ of $V$
admits a coloring $c$ of $G$  proper to $(G,\sigma)$.
 By the definition of indices $a(i)$, $b(i)$, $i_{\tt red}$ and $i_{\tt blue}$ for a coloring $c$ of $G$,
 we easily observe the following property.

\begin{lemma}\label{le:color}
For a graph $G=(V,E)$ with $n\geq 2$, an ordering $\sigma=v_1,v_2,\ldots,v_n$ of $V$
and a coloring $c$ of $G$  proper to $(G,\sigma)$,
 the following holds.
 \begin{enumerate}
\item[{\rm (i)}]
Every two indices $i$ and $j$ with   $1\leq i<j\leq n$ satisfy
$a(j)\leq a(i)$ and $b(j)\leq b(i)$;
\item[{\rm (ii)}]
 It holds that $i_{\tt red}+1 \leq i_{\tt blue}-1$,
  $(i,j)\in {\tt red}$ for $i<j\leq i_{\tt red}$,
 $(i,j)\in {\tt blue}$ for $i_{\tt blue}\leq i<j$, and
 $(i,j)\in {\tt green}$ for $i_{\tt red}<i<j<i_{\tt blue}$.
\item[{\rm (iii)}]
Each index $p\in [1, i_{\tt red}]$ 
satisfies
  $ p+2\leq  i_{\tt red} +2\leq a(i_{\tt red}) \leq a(p)$; and
\item[{\rm (iv)}]
Each index $q\in [ i_{\tt blue},n]$ 
satisfies
   $ b(q)\leq b(i_{\tt blue})\leq i_{\tt blue}-2\leq q-2$.
\end{enumerate}
\end{lemma}
 \noindent {\bf Proof.}
 (i) Let $1\leq i<j\leq n$.
 If $a(i)<a(j)$, then $(i,a(i))\in {\tt green}$ while $(j,a(i))\in {\tt red}$,
 contradicting that coloring $c$ is proper to $\sigma$.
 If $b(i)<b(j)$, then $(b(j),j)\in {\tt green}$ while $(b(j),i)\in {\tt blue}$,
 contradicting that coloring $c$ is proper to $\sigma$.

 (ii) By definition, $i_{\tt red}\leq n-1$  and $i_{\tt blue}\geq 2$.
 Then if $i_{\tt red}=0$ or $i_{\tt blue}=n+1$, then $ i_{\tt red}+2\leq i_{\tt blue}$ holds.
 When $E_{\tt red},E_{\tt blue}\neq\emptyset$,
 the index $i_{\tt red}$ is the largest $i$ with $(i, i+1)\in {\tt red}$,
 while $i_{\tt red}$ is the smallest $i$ with  $(i-1,i)\in {\tt blue}$.
 See Fig.~\ref{fi:degree_cut} for an illustration of a star $T_V$.
 The former  implies that
 $(j, i_{\tt red}+1)\in {\tt red}$ for all $j\leq i_{\tt red}$.
Hence if $(i-1,i)\in {\tt blue}$ then $i_{\tt red}+1\leq i-1$,
indicating that  $ i_{\tt red}+1\leq i_{\tt blue}-1$, as required.

For every $i_{\tt red}<i<j$, it holds  $(i,j)\not\in {\tt red}$
by the  definition of $i_{\tt red}$,
and
for every $i<j<i_{\tt blue}$, it holds  $(i,j)\not\in {\tt blue}$
 by the  definition of $i_{\tt blue}$.
This means that $(i,j)\in {\tt red}$ for $i<j\leq i_{\tt red}$,
 $(i,j)\in {\tt blue}$ for $i_{\tt blue}\leq i<j$, and
 $(i,j)\in {\tt green}$ for $i_{\tt red}<i<j<i_{\tt blue}$.

 (iii) By choice of $p$, it holds that $ p+2\leq  i_{\tt red} +2$.
 By this and (i), it holds that $a(i_{\tt red}) \leq a(p)$.
 By definition of $i_{\tt red}$, it holds that
 $i_{\tt red} +2\leq a(i_{\tt red})$.

 (iv) Analogous with (iii).
  \qed
\bigskip

\begin{figure}[htbp] \begin{center}
\includegraphics[scale=0.22]{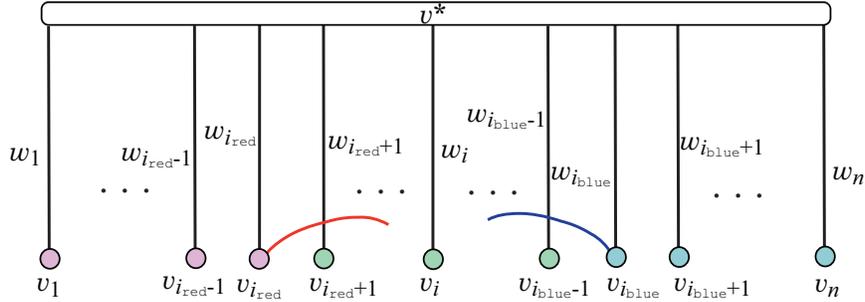} \end{center}
\caption{Illustration of a star $T_V$ with  weights $w_i=w(v^*v_i)$ of edge $v^*v_i \in E(T)$. }
\label{fi:degree_cut}
\end{figure}

  We determine a real value to each weight $w_i$,
  where the value may be negative.
  Recall that we can convert the weights of leaf edges into positive values if necessary
  without changing the tree in a PCR.

  For some $p,q\in [1,n]$ with $p< q$, suppose that
  weights $w_i$ with $p+1\leq i\leq q-1$ and bounds $ d_{\min}$ and $ d_{\max}$ have been determined so that\\
\hspace*{1mm} (a)    $ w_{p+1}\leq w_{p+2}\leq \cdots \leq w_{q-1} $; \\
\hspace*{1mm} (b) $ w_{j} < w_{j+1}$ for   $\{v_j, v_{j+1}\}\not\in M_G$ with $j\in [p+1, q-2]$; \\
\hspace*{1mm} (c)   $w_j + w_k > d_{\min}$ for  $(j,k)\in {\tt green}$ with $p+1\leq j<k\leq q-1$,\\
\hspace*{3mm}  $w_j + w_k < d_{\min}$ for $(j,k)\in {\tt red}$ with $p+1\leq j<k\leq q-1$,\\
\hspace*{3mm}  $w_h + w_j  < d_{\max}$ for $(h,j)\in  {\tt green}$ with $p+1\leq h<j\leq q-1$,\\
\hspace*{3mm} $w_h + w_j  > d_{\max}$ for $(h,j) \in {\tt blue}$ with $p+1\leq h<j\leq q-1$.\\
Hence for distinct $i,i'\in [p+1, q-1]$, it holds
  $ d_{\min}>w_i + w_{i'}$ if   $(i, i')\in {\tt red}$;
   $ d_{\min}<w_i + w_{i'}< d_{\max}$ if   $(i, i')\in {\tt green}$; and
  $  w_i + w_{i'} >d_{\max}$ if   $(i, i')\in {\tt blue}$. \\

 With the next lemma, we start with
  weights $w_i$ with $p+1\leq i\leq q-1$ and bounds $ d_{\min}$ and $ d_{\max}$
  for $p=i_{\tt red}$ and $q=i_{\tt blue}$.

\begin{lemma}\label{le:middle}
For a graph $G=(V,E)$ with $n\geq 2$ and an ordering $\sigma=v_1,v_2,\ldots,v_n$ of $V$,
let $c$ be  a coloring $c$ of $G$  proper to $(G,\sigma)$. 
 For $p=i_{\tt red}$ and $q=i_{\tt blue}$,
 there is  a set
 $\{ w_{p+1}, w_{p+2},\ldots,w_{q-1}, d_{\min}, d_{\max}\}$ of weights and bounds
 that satisfies  condition (a)-(c).
\end{lemma}
 \noindent {\bf Proof.}
 Since $(i,j)\in {\tt green}$ for all $i_{\tt red}<i<j<i_{\tt blue}$
by Lemma~\ref{le:color}(i), we  can easily find required weights $w_i$ for all $i$ with $i_{\tt red}<i<i_{\tt blue}$
so that   $d_{\min}<w_i+w_{i'}<d_{\max}$ for all $i,i'$ with $i_{\tt red}<i<i'<i_{\tt blue}$.
For example, such weights $w_i$ and bounds $d_{\min}$ and  $d_{\max}$ can be obtained
as follows.

\noindent
~~~~ $d_{\min}:=0$; 
 $w_{i_{\tt red}+1}:=1$;\\
~~~~ {\bf for} $i=i_{\tt red}+2,i_{\tt red}+3,\ldots, i_{\tt blue}-1$ {\bf do} \\
~~~~ {\bf if} $\{ v_i, v_{i+1} \}\in M_G$ {\bf then}\\
~~~~  ~~ $w_{i+1} : = w_{i }$ \\
~~~~  {\bf else} \\
~~~~  ~~  $w_i : = w_{i+1} +1$ \\
~~~~  {\bf end if};\\
~~~~    $d_{\max}:=2w_{i_{\tt blue}-1} +1$.\\
 \qed
\bigskip


 Then without changing the determined values to $ w_{p+1},w_{p+2}, \ldots ,w_{q-1} $,
 we determine $w_p$ or $w_q$ so that the above condition holds
 for $(p:=p-1,q)$ or $(p,q:=q+1)$.

 \bigskip


\begin{figure}[htbp] \begin{center}
\includegraphics[scale=0.16]{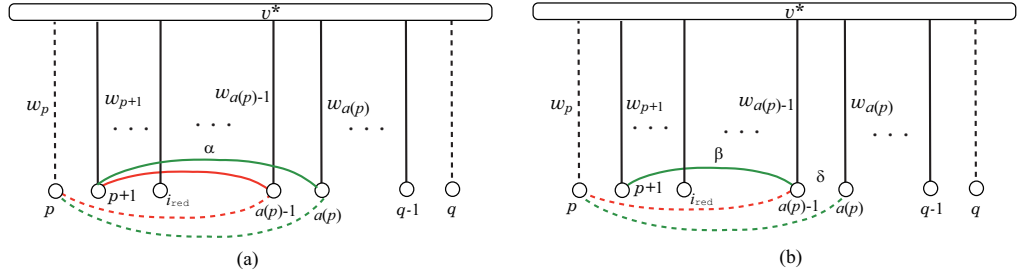} \end{center}
\caption{Illustration of a process of determining weight $w_p$,
where vertex $v_i$ is indicated with its index $i$:
(a) Case (ii) where
 $\{v_p, v_{p+1} \}\not\in M_G$ and $a(p+1)=a(p)\leq q-1$;
 (b) Case (iii) where
 $\{v_p, v_{p+1} \}\not\in M_G$ and $a(p+1)<a(p)\leq q-1$. }
\label{fi:degree_cut}
\end{figure}

 We can determine values to weights $w_{i_{\tt red}}, w_{i_{\tt red}-1}, \ldots,w_1$
 in this order as follows.

\begin{lemma}\label{le:low-part}
For a graph $G=(V,E)$ with $n\geq 2$ and an ordering $\sigma=v_1,v_2,\ldots,v_n$ of $V$,
let $c$ be  a coloring $c$ of $G$  proper to $(G,\sigma)$. 
 For $p\leq i_{\tt red}$ and $q\geq i_{\tt blue}$, assume that a set
 $\{ w_{p+1}, w_{p+2},\ldots,w_{q-1}, $ $d_{\min}, d_{\max}\}$ of weights and bounds
  satisfies conditions (a)-(c).
  When $p\geq 1$, let weight $w_p$ be determined such that
\[w_p =
 \left\{ \begin{array}{cl}
 w_{p+1}   &  \mbox{ if $\{v_p, v_{p+1} \}\in M_G$, }   \\
  w_{p+1} - \alpha/2 &   \mbox{ if $\{v_p, v_{p+1} \}\not\in M_G$ and $a(p+1)=a(p)\leq q-1$, }\\
  & \mbox{~~~ where  $\alpha=w_{p+1}+w_{a(p)} -d_{\min}~(>0)$,  }   \\
  w_{p+1} -\beta-\delta/2   &  \mbox{ if $\{v_p, v_{p+1} \}\not\in M_G$ and $a(p+1)<a(p)\leq q-1$, }\\
  & \mbox{~~~ where  $\beta= w_{p+1} + w_{a(p)-1} - d_{\min}~(>0)$ and} \\
  & \mbox{~~~~~~$\delta =w_{a(p)}-w_{a(p)-1}~(>0)$, }   \\
   \min\{ w_{p+1}, d_{\min}\!-\!w_{q-1}\}\!-\!1 & \mbox{ otherwise, i.e., $\{v_p, v_{p+1} \}\not\in M_G$ and $q\leq a(p)$.}
 \end{array} \right.
\]
Then   the set
 $\{ w_{p}, w_{p+1}, w_{p+2},\ldots,w_{q-1}, d_{\min},d_{\max}\}$ of weights and bounds
  satisfies conditions (a)-(c) for $p'=p-1$ and $q$.
\end{lemma}
 \noindent {\bf Proof.} We distinguish the four cases.

  (i) $\{v_p, v_{p+1} \}\in M_G$:
  Since $\{v_p, v_{p+1} \} \in M_G$ and $w_p =w_{p+1}$,
  we see that for each $i\in [p+2,q]$ with $(p,i)\in {\tt green}$ (resp.,  $(p,i)\in {\tt red}$, ${\tt blue}$)
 $d_{\min}<w_p+w_i=w_{p+1}+w_i<d_{\max}$ (resp.,
 $w_p+w_i=w_{p+1}+w_i<d_{\min}$ and $w_p+w_i=w_{p+1}+w_i>d_{\max}$).
      To show that conditions (a)-(c) hold for $(p':=p-1,q)$,
       it suffices to show that $(p,p+1)\in {\tt red}$ and $w_p+w_{p+1}<d_{\min}$.
 If $p=i_{\tt red}$ then $i_{\tt red}$ would be larger than the current $p$
 since $\{v_p, v_{p+1} \} \in M_G$.
 Hence  $p<i_{\tt red}$, which implies that
 $p+2\leq {\tt red}+1$ and $(p,p+1),(p+1,p+2)\in {\tt red}$.
 Since $  {\tt red}+2\leq i_{blue} $
 by Lemma~\ref{le:color}(ii),
 Then  $w_{p+1}$ and $w_{p+2}$ have been determined
 so that $w_{p+1}+w_{p+2}<d_{\min}$ since condition  (c) holds.
 Therefore  $w_{p}+w_{p+1}\leq w_{p+1}+w_{p+2}<d_{\min}$ since condition (a) holds.

 (ii)   $\{v_p, v_{p+1} \}\not\in M_G$ and $a(p+1)=a(p)\leq q-1$:
 Fig.~\ref{fi:degree_cut}(a) illustrates the case where
 $\{v_p, v_{p+1} \}\not\in M_G$ and $a(p+1)=a(p)\leq q-1$.
 Since  condition (a) holds and $a(p)\leq q-1$, it holds that
   $w_{p+1}\leq \cdots \leq w_{a(q)-1}\leq w_{a(q)}\leq w_{q-1}$.
 Since $a(p+1)=a(p)$,  we see that $(p,a(p)-1),(p+1,a(p)-1)\in {\tt red}$ and $(p,a(a)), (p+1,a(p))\in {\tt green}$.
  Since condition (c) holds for $(p+1,a(p))\in {\tt green}$,
   we see that   $\alpha= w_{p+1} + w_{a(p)} - d_{\min}$ is  positive.
      To show that conditions (a)-(c) hold for $(p':=p-1,q)$,
      it suffices to show that $w_p<w_{p+1}$; $w_p+w_{a(p)}>d_{\min}$; and $w_p+w_{a(p)-1}<d_{\min}$.
      Since $\alpha>0$, we have $w_p=w_{p+1} -\alpha/2 <w_{p+1}$.
  We next see that   
    $w_p+w_{a(p)}=(w_{p+1}-\alpha/2)  + w_{a(p)} =d_{\min}+\alpha/2> d_{\min}$, as required.
      Finally we observe that  $w_p+w_{a(p)-1} < w_{p+1}+ w_{a(p)-1}<d_{\min}$ since
       condition (c) holds for $(p+1,a(p)-1)\in {\tt red}$.

  (iii)  $\{v_p, v_{p+1} \}\not\in M_G$ and $a(p+1)<a(p)\leq q-1$:
 Fig.~\ref{fi:degree_cut}(b) illustrates the case where
 $\{v_p, v_{p+1} \}\not\in M_G$ and $a(p+1)<a(p)\leq q-1$
    As in (ii), we see  that
   $w_{p+1}\leq \cdots \leq w_{a(q)-1}\leq w_{a(q)}\leq w_{q-1}$.
   Since $a(p+1)<a(p)$,  we see that $(p,a(p)-1)\in {\tt red}$ and $(p+1,a(p)-1)\in {\tt green}$.
   This means that $\{v_{a(p)-1}, v_{a(p)}\}\not\in M_G$, where $w_{a(q)-1}< w_{a(q)}$
   by condition (b) and $\delta =w_{a(p)}-w_{a(p)-1}$ is positive.
  Since condition (c) holds for $(p+1,a(p)-1)\in {\tt green}$,
      $\beta= w_{p+1} + w_{a(p)-1} - d_{\min}$ is  positive.
      To show that conditions (a)-(c) hold for $(p':=p-1,q)$,
      it suffices to show that $w_p<w_{p+1}$; $w_p+w_{a(p)}>d_{\min}$; and $w_p+w_{a(p)-1}<d_{\min}$.
      Since $\beta,\delta>0$, we have $w_p=w_{p+1} -\beta-\delta/2 <w_{p+1}$.
  We next see that     $w_p+w_{a(p)}=(w_{p+1}-\beta-\delta/2)  +(w_{a(p)-1}+\delta)
                              =  d_{\min}+\delta/2> d_{\min}$, as required.
      Finally we observe that
       $w_p+w_{a(p)-1} =(w_{p+1}-\beta-\delta/2) +w_{a(p)-1}=d_{\min}-\delta/2<d_{\min}$, as required.

  (iv)   $\{v_p, v_{p+1} \}\not\in M_G$ and $q\leq a(p)$:
  Since $q\leq a(p)$, $(p,i)\in {\tt red}$ for all $i\in [p+1,q-1]$.
  As in (ii), we see  that
   $w_{p+1}\leq \cdots \leq w_{q-1}$.
      To show that conditions (a)-(c) hold for $(p':=p-1,q)$,
      it suffices to show that $w_p<w_{p+1}$; and  $w_p+w_{q-1}<d_{\min}$.
  Obviously $w_p=\min\{ w_{p+1}, d_{\min}- w_{q-1}\}-1\leq w_{p+1}-1<w_{p+1}$, as required.
   Finally we observe that  $w_p+w_{q-1} \leq (d_{\min} - w_{q-1}- 1)+w_{q-1}=d_{\min}-1 <d_{\min}$, as required.
 \qed
\bigskip

Analogously with the process of computing   weights $w_{i_{\tt red}}, w_{i_{\tt red}-1}, \ldots,w_1$
 by  Lemma~\ref{le:low-part},
 we can choose  weights $w_{i_{\tt blue}}, w_{i_{\tt blue}+1}, \ldots,w_n$
 in this order as follows.

\begin{lemma}\label{le:upper-part}
For a graph $G=(V,E)$ with $n\geq 2$ and an ordering $\sigma=v_1,v_2,\ldots,v_n$ of $V$,
let $c$ be  a coloring $c$ of $G$  proper to $(G,\sigma)$. 
 For $p\leq i_{\tt red}$ and $q\geq i_{\tt blue}$, assume that a set
 $\{ w_{p+1}, w_{p+2},\ldots,w_{q-1},$ $ d_{\min}, d_{\max}\}$ of weights and bounds
  satisfies conditions (a)-(c).
  When $q\leq n$, let weight $w_q$ be determined such that
\[w_q =
 \left\{ \begin{array}{cl}
 w_{q-1}   &  \mbox{ if $\{ v_{q-1}, v_q \}\in M_G$, }   \\
  w_{q-1} + \alpha/2 &   \mbox{ if $\{v_{q-1}, v_q \}\not\in M_G$ and $p+1\leq b(q)=b(q-1)$, }\\
  & \mbox{~~~ where  $\alpha=d_{\max}-(w_{b(p)}+w_{ q-1})~(>0)$,  }   \\
  w_{q-1} +\beta+\delta/2   &  \mbox{ if $\{v_{q-1}, v_q \}\not\in M_G$ and $p+1\leq b(q)<b(q-1)$, }\\
  & \mbox{~~~ where  $\beta= d_{\max}-( w_{b(p)+1}+w_{q-1})~(>0)$ and} \\
  & \mbox{~~~~~~$\delta =w_{b(q)+1}-w_{b(q)}~(>0)$, }   \\
   \max\{ w_{q-1}, d_{\max}\!-\!w_{q-1}\}\!+\!1 & \mbox{ otherwise, i.e., $\{v_{q-1}, v_q \}\not\in M_G$ and $b(q)\leq p$.}
 \end{array} \right.
\]
  Then   the set
 $\{  w_{p+1}, w_{p+2},\ldots,w_{q-1}, w_{q}, d_{\min},d_{\max}\}$ of weights and bounds
  satisfies conditions (a)-(c) for $p$ and $q':=q+1$.
\end{lemma}
 \noindent {\bf Proof.}
 Analogously with the proof of Lemma~\ref{le:low-part}.
 \qed
\bigskip

We are ready to prove Lemma~\ref{le:compute_weight}.
Given  a graph $G=(V,E)$ with 
  an gap-free ordering $\sigma=v_1,v_2,\ldots,v_n$  of $V$,
   a coloring $c$ of $G$  proper to $(G,\sigma)$ can be obtained in $O(n^2)$ time
  by Lemma~\ref{le:proper_color}. 
For the coloring $c$, 
we first  comput $M_G$, indices $a(i),b(i)$,
 $i=1,2,\ldots,n$ and $i_{\tt red}$ and $i_{\tt blue}$ for 
in $O(n^3)$ time.
Based on these, 
we can determine weights $w_i$ with  $i_{\tt red}<i<i_{\tt blue}$ in $O(n)$ time
by the method in the proof of Lemma~\ref{le:middle}.
Finally we determine weights $w_i$ with $1\leq i\leq i_{\tt red}$ and
$i_{\tt blue}\leq i\leq n$ in $O(n)$ time by
Lemmas~\ref{le:low-part} and \ref{le:upper-part}, respectively.
This gives  a PCR $(T_V,w,d_{\min},d_{\max})$ of $G$ such that
$w_1\leq w_2\leq\cdots\leq w_n$  in $O(n^3)$ time. 
This proves Lemma~\ref{le:compute_weight}.

\section{Proof of Lemma~\ref{le:structure1}}

\textbf{Lemma~\ref{le:structure1}}
\emph{For a graph $G=(V,E)$ with
  a gap-free  ordering $\sigma= v_1,v_2,\ldots,v_n$  of  $V$ and
    a   coloring $c$  proper to $\sigma$,
    let   $V_1=\{v_i\mid 1\leq i\leq i_{\tt red}\}$,
     $V_2=\{v_i\mid i_{\tt blue}\leq i\leq n\}$, and
     $V^*=\{v_i \mid i_{\tt red}-1 \leq i\leq  i_{\tt blue}+1\}$.
Then
\begin{enumerate}
\item[{\rm (i)}]
If two edges $v_i v_j$ and $v_{i'}  v_{j'}$
 with $i<j$ and $i'<j'$   cross $($i.e.,   $i<i'<j<j'$ or $i'<i<j'<j)$,
 then they belong to the same component  of $G$;
\item[{\rm (ii)}]
 It holds
 $i_{\tt red}+1 \leq i_{\tt blue}-1$.
 The graph $G[V^*]$ is a complete graph,
 and $G-V^*$ is a bipartite graph between vertex sets  $V_1$ and $V_2$; 
 and
\item[{\rm (iii)}]
Every two vertices $v_i,v_j\in V_1\cap N_G(V^*)$ with $i<j$
satisfy  $v_{i_{\tt blue}-1}\in N_G(v_i)\cap V^*\subseteq N_G(v_j)\cap V^*
\subseteq V^*\setminus \{v_{i_{\tt red}+1}\}$; and \\
Every two vertices $v_i,v_j\in V_2\cap N_G(V^*)$ with $i<j$
satisfy   $v_{i_{\tt red}+1}\in N_G(v_j)\cap V^*\subseteq N_G(v_i)\cap V^*
\subseteq V^*\setminus \{v_{i_{\tt blue}-1}\}$.
\end{enumerate}}
\medskip

\noindent {\bf Proof.}
(i) Let edges $e=v_iv_j,e'=v_{i'}v_{j'}\in E$ cross, where
$i<i'<j<j'$.
If $e$ and $e'$ belong 
to
  different
components   of $G$,   then $\{v_{i'},v_j\}$ would a gap with respect  to
edges $e$ and $e'$.

(ii)
Immediate from Lemma~\ref{le:color}(ii).
 
(iii) We show that
every two vertices $v_i,v_j\in V_1\cap N_G(V^*)$ with $i<j$
satisfy  $v_{i_{\tt blue}-1}\in N_G(v_i)\cap V^*\subseteq N_G(v_j)\cap V^*
\subseteq V^*\setminus \{v_{i_{\tt red}+1}\}$ (the other case can be treated
symmetrically).
For this, it suffices to show that,
for any vertex $v_i\in V_1$ with $N_G(v_i)\cap V^*\neq\emptyset$, \\
~~ (a) it holds $v_{i_{\tt blue}-1}\in N_G(v_i)$; and\\
~~ (b) for any vertex $v\in N_G(v_j)\cap V^*$ with $v_j\in V_1$ and $i<j$, \\
~~~~~ it holds that $v\in N_G(v_i)$. \\
In (a), otherwise $\{v_i,v_{i_{\tt blue}-1}\}$ would be a gap
with respect to edges $v_iv$ and $vv_{i_{\tt blue}-1}$ for any vertex $v\in N_G(v_i)\cap V^*$.
In (b), otherwise $\{v_j,v \}$ would be a gap
with respect to edges $v_iv$ and $v_jv_{i_{\tt blue}-1}$.
%
\qed \bigskip

\section{Proof of Lemma~\ref{le:bipartite_case}}

\textbf{Lemma~\ref{le:bipartite_case}}
\emph{Let $G=(V_1,V_2,E)$ be a connected  bipartite  graph with $|E|\geq 1$.
Then  family $\mathcal{S}_i$ is separator-free for each $i=1,2$,
and $G$ has a gap-free ordering of $V$ 
 if and only if for each $i=1,2$, family $\mathcal{S}_i$ admits a contiguous ordering $\sigma_i$ of $V_i$.
 For any contiguous ordering $\sigma_i$ of $V_i$, $i=1,2$,
 one of orderings  $(\sigma_1,\sigma_2)$ and  $(\sigma_1, \overline{\sigma_2})$
  of $V$  is  a gap-free ordering to $G$.}
\medskip

\noindent {\bf Proof.}
Since $G$ is connected, we see that, for each $i=1,2$, the family  $\mathcal{S}_i$ is separation-free.

The  only if part:
Let $v_1,v_2,\ldots,v_k,v_{k+1},\ldots,v_n$ be a gap-free ordering of $V$ to $G$,
where $V_1=\{v_1,v_2,\ldots,v_k\}$ and $V_2=\{v_{k+1},\ldots,v_n\}$.
Since there is no gap, for each vertex $v\in V_2$, the neighbors in $N_G(v)$ appear
consecutively as a subsequence of $v_1,v_2,\ldots,v_k$.
Also for any two vertices $u,v\in V_2$ such that  $N_G(u)$ is a proper subset of $N_G(v)$,
the subsequence $v_i,v_{i+1},\ldots,v_{j}$ for $N_G(u)$ must
 contain the first vertex or the last vertex in the subsequence $v_{h},v_{h+1},\ldots,v_p$ for $N_G(v)$.
This is because otherwise $v_h,v_p\not\in N_G(u)$ would imply that $\{u,v_h\}$ (or $\{u,v_p\}$)
is a gap with respect to edges $e_1=uv_i$ and $e_1=vv_h$ (or $e_1=uv_j$ and $e_1=vv_p$).
Therefore $v_1,v_2,\ldots,v_k$  is a contiguous ordering   of $V_1$ to
$\mathcal{S}_1$.
Analogously with $V_2$ and $\mathcal{S}_2$.

The if part:
Assume that for each $i=1,2$, the family  $\mathcal{S}_i$ has a   contiguous ordering $\sigma_i$ of $V_i$.
Note that any set $X\in \mathcal{X}_{\mathcal{S}_i}$, $i=1,2$ is either contained in $N_G(v)$
or disjoint with $N_G(v)$ for each vertex $v\in V_j$, $j\neq i$.
By Theorem~\ref{th:contiguous} 
applied to $\mathcal{S}_i$,
the vertices in each maximal set $X\in \mathcal{X}_{\mathcal{S}_i}$
appear consecutively in any contiguous ordering of $V(\mathcal{S}_i)$.
Also a  contiguous ordering of $V(\mathcal{S}_i)$
is unique up to reversal and choice of an ordering
of each set $X\in \mathcal{X}_{\mathcal{S}_i}$.
This means that an ordering
$\sigma=(\sigma_1,\sigma_2)$ or  $(\sigma_1, \overline{\sigma_2})$  of $V$
 is gap-free if and only if
any ordering obtained from $\sigma$ by
changing an ordering of vertices in each set
$X\in \mathcal{X}_{\mathcal{S}_1}\cup \mathcal{X}_{\mathcal{S}_2}$.
Therefore, to see if $G$ admits a gap-free ordering of $V$,
we only need to check if at least one of
$(\sigma_1,\sigma_2)$ and  $(\sigma_1, \overline{\sigma_2})$ is gap-free in $G$.
\qed \bigskip

\section{Proof of Lemma~\ref{le:structure2}}

\textbf{Lemma~\ref{le:structure2}}
\emph{For a connected non-bipartite  graph  $G=(V,E)$  with $V^{\tt t}\neq \emptyset$,
and let $v^*_1,v^*_2$ be two adjacent vertices in $V^{\tt t}$.
Let $V^*=\{v^*_1,v^*_2\}\cup(N_G(v^*_1)\cap N_G(v^*_2))$,
$V'_1=N_G(v^*_2)\setminus V^*$, and
$V'_2=N_G(v^*_1)\setminus V^*$.
Assume that $G$ has a gap-free ordering $\sigma$ of $V$ and a proper coloring $c$ to $\sigma$
 such that
$v^*_1=v_{i_{\tt red}+1}$, $v^*_2=v_{i_{\tt blue}-1}$.
Then:
\begin{enumerate}
\item[{\rm (i)}]
A maximal clique $K_{v^*_1,v^*_2}$ of $G$ that contains edge $v^*_1,v^*_2$
  is uniquely given as
  $G[V^*]$.
The graph $G[V^*]$ is the core of  the ordering $\sigma$,
 and $G-V^*$ is a bipartite graph $(V_1,V_2,E')$; and
\item[{\rm (ii)}]
Let $\mathcal{S}_i$ be the family $\{N_G(v)\mid v\in V_j\}$ for $\{i,j\}=\{1,2\}$,
 and  $\mathcal{S}=\mathcal{S}_1\cup \mathcal{S}_2\cup\{ V^* \}$.
Then
$\mathcal{S}$ is a separator-free family 
that admits a contiguous ordering $\sigma$ of $V$,
  and any  contiguous ordering $\sigma$ of $V$ is  a gap-free ordering to $G$.
\end{enumerate}}
\medskip

\noindent {\bf Proof.}
(i)
By Lemma~\ref{le:structure11}(ii), we see that
$\{v_i\mid  i_{\tt red}+1 <i< i_{\tt blue}-1   \}\subseteq  N_G(v_{i_{\tt red}+1})\cap N_G(v_{i_{\tt blue}-1})$.
On the other hand, by Lemma~\ref{le:structure11}(ii),
vertex $v_{i_{\tt red}+1}$ (resp., $v_{i_{\tt blue}-1}$) is not adjacent to any vertex in $V_2$
(resp., $V_1$) in $G$.
Hence
$\{ v_{i_{\tt red}+1}, v_{i_{\tt blue}-1}\}\cup (N_G(v_{i_{\tt red}+1})\cap N_G(v_{i_{\tt blue}-1}))$
induces uniquely a maximal clique that contains $v_{i_{\tt red}+1}$ and $v_{i_{\tt blue}-1}$.
Hence the clique is the core of the gap-free ordering of $V$.
By Lemma~\ref{le:structure1}(ii),
   $G-V^*$ is a bipartite graph $(V_1,V_2,E')$.

(ii) 
Since $G$ is connected, we see that   $\mathcal{S}_i$ is separation-free.
First we prove that  $\mathcal{S}$ admits
a contiguous ordering of $V$.
  Any set $S\in \mathcal{S}$ with $S\cap V^*\neq\emptyset$ satisfies
  one of the following:\\
~~- $v_{i_{\tt red}+1}\not\in S$ and $v_{i_{\tt blue}-1}\in S\in \mathcal{S}_2$;\\
 ~~-  $v_{i_{\tt blue}-1}\not\in S$ and $v_{i_{\tt red}+1}\in S\in \mathcal{S}_1$; and\\
~~- $v_{i_{\tt red}+1}, v_{i_{\tt blue}-1} \in S=V^*$. \\
This means that  any two sets $S,S'\in \mathcal{S}$ with $S\subseteq S'$ belong
  to the same family $\mathcal{S}_i$.
  Hence any gap-free ordering $\sigma$ of $V$ to $G$
  is a contiguous ordering of $V$ to $\mathcal{S}$, as discussed
  in the proof of the only if part of Lemma~\ref{le:bipartite_case}.

Next we prove that any contiguous ordering $\sigma$ of $V$ to  $\mathcal{S}$
is a gap-free ordering of $V$ in $G$.
Since $G$ is connected, each $\mathcal{S}_i$ is separator-free in $V(\mathcal{S}_i)$.
We see that   $\mathcal{S}=\mathcal{S}_1\cup \mathcal{S}_2\cup\{ V^* \}$
  is separator-free in $V$ even if $V(\mathcal{S}_1)\cap V(\mathcal{S}_2)\neq\emptyset$.
Note that any set $X\in \mathcal{X}_{\mathcal{S}}$  is either contained in $N_G(v)$
or disjoint with $N_G(v)$ for each vertex $v\in V_j$, $j\neq i$.
By Theorem~\ref{th:contiguous} 
applied to $\mathcal{S}$,
the vertices in each maximal set $X\in \mathcal{X}_{\mathcal{S}}$
appear consecutively in any contiguous ordering of $V(\mathcal{S})$.
Also a  contiguous ordering of $V(\mathcal{S})$
is unique up to reversal and choice of an ordering
of each set $X\in \mathcal{X}_{\mathcal{S}}$.
This means that an ordering $\sigma$  of $V$
 is gap-free if and only if
any ordering obtained from $\sigma$ by
changing an ordering of vertices in each set
$X\in \mathcal{X}_{\mathcal{S}}$.
Therefore any contiguous ordering $\sigma$ of $V$ to  $\mathcal{S}$
is a gap-free ordering of $V$ in $G$.
\qed \bigskip

\end{document}